\documentclass[preprint,12pt,authoryear]{elsarticle}

\usepackage{amssymb}
\usepackage{amsmath}

\usepackage{amsfonts}
\usepackage{esint}
\usepackage{graphicx}
\usepackage{subfigure}
\usepackage{color}
\usepackage{epstopdf}
\usepackage{tikz}
\usepackage{pdflscape}
\usepackage{mathtools}
\usepackage{rotating}

\graphicspath{{figures/}} 

\journal{International Journal of Engineering Science}

\begin{document}
	
\begin{frontmatter}
		
		
		
\title{{Torsional vibration of a coupled cylinder}
} 

%
\author[RZ]{Igor Istenes}
\ead[url]{igor.istenes@roez.sk}

\author[AU]{Daniel Peck}
\ead[url]{dtp@aber.ac.uk}

\author[OD]{Yuriy Protserov}
\ead[url]{protserov@onu.edu.ua}

\author[AU,LD]{Natalya Vaysfeld}
\ead[url]{natalya.vaysfeld@kcl.ac.uk}

\author[OD]{Zinaida Zhuravlova}
\ead[url]{z.zhuravlova@onu.edu.ua}

\affiliation[RZ]{
		organization={Roez R{\&}D},
		city={Bratislava},
		postcode={811 04},
		country={Slovakia}}

\affiliation[AU]{
		organization={Department of Mathematics},
	    addressline={Aberystwyth University},
		city={Aberystwyth},
		postcode={SY23 3BZ},
		state={Wales},
		country={United Kingdom}}

\affiliation[OD]{
		organization={Odesa I.I. Mechnikov National University},
		addressline={Faculty of Mathematics, Physics and Information Technologies, str. Dvoryanska, 2},
		city={Odesa},
		postcode={65082},
		country={Ukraine}}

\affiliation[LD]{
		organisation={King's College},
		addressline={London},
		city={London},
		postcode={S2.35},
		country={United Kingdom}}

\begin{abstract}
	The torsion loading of a coupled cylinder, {comprising distinct upper and lower cylindrical sections potentially made of different materials}, is considered. The bottom of the cylinder is fixed in place, and induces the cylinder {vibration}. The torsion is applied via an arbitrary loading on the upper face. Three forms of coupling condition between the upper and lower cylinders are outlined: ideal, soft (weak), and rigid (hard/ stiff) contact. The resulting displacements and tangential stresses are obtained using the finite Hankel transform, and a Green's function representation of the displacement. Numerical results are provided, and the impact of the differing coupling conditions investigated for a range of {cylinder geometries, material properties and vibration rates. The resonance frequencies of the coupled cylinder are determined.} A method for using the coupled cylinder model to approximate the displacement of a cylinder containing a damaged region via a weak interfacial layer is outlined. The properties of the weak interface layer needed for this approximation are determined, and the advantages of its use in non-destructive testing are discussed.
\end{abstract}


\begin{highlights}
	\item The displacement and stress for a {vibrating} coupled cylinder under torsion are obtained.
	\item The impact of differing contact conditions: ideal, soft or rigid, is investigated. 
	\item {The resonance frequencies of the cylinder are determined.}
	\item A method for approximating cylinder damage using a weak interfacial layer is outlined, with application to non-destructive testing.
\end{highlights}


\end{frontmatter}




\section{Introduction}

{Torsion problems involve applying torque to an object, causing it to deform through twisting. Understanding torsion in layered or coupled elastic cylinders - both solid and hollow - is fundamental for analysing composite structures in engineering. Such studies underpin applications in aerospace and civil engineering (e.g. turbines), biomedical implants (e.g. spinal or joint components), and soft robotics (e.g. actuators), where layered configurations offer tailored mechanical properties.}

{Layered construction introduces complex interactions due to the unique mechanical responses it induces combined with the constraints of finite dimensions, requiring advanced modelling techniques. Classical solutions, such as those based on Saint-Venant’s principle, have been extended for layered systems, while computational approaches such as finite element (FEA) and boundary element methods (BEM) are now standard for handling multi-layered geometries. Recent work also explores machine learning for parameter optimisation and scenario simulation, signalling a shift toward hybrid analytical–data-driven methods.}

{Significant progress has been made in modelling torsion in finite cylinders. For example, \cite{Protserov2017} solved the multilayered cylinder problem with a fixed base, while \cite{WANG199225} addressed the elastodynamic case without a fixed bottom. \cite{LI2013113} analysed interface cracks in bi-layered composites using Fourier transforms and singular integral equations, and \cite{Zingerman2023} derived exact solutions for large deformations in multilayer rods composed of incompressible isotropic nonlinear elastic materials. Nonlinear torsion in rubber cylinders was studied by \cite{FALOPE2025104254}. Investigations into cracks within solid or hollow cylinders were presented by \cite{Vaysfeld2017}, and extended to the case of a hollow cylinder under torsion by \cite{ZHURAVLOVA2024103976}. Torsional wave dispersion in pre-stressed bimaterial cylinders was examined by \cite{AKBAROV20114519}, later incorporating imperfect (shear-spring type) contact conditions \cite{AKBAROV2025}. Classical solutions for reinforced tubes were given by \cite{KUO19731553}, while recent studies have addressed porous cylindrical shells \cite{ESFANDYARI2025108194} and radially stratified spherical shells \cite{AKHMEDOV2009296}, with the latter establishing that the type of contact between layers - whether ideal, soft, or rigid (hard) - greatly influences how stresses and deformations propagate through the structure.}

{The role of interfacial conditions - ideal, soft, or rigid - have also been found to either enhance or reduce the cylinder’s overall mechanical performance, depending on the application  \cite{Mishuris2006SteadystateMO}. Ideal contact assumes perfect bonding, soft contact allows limited slip via a compliant layer, and rigid contact restricts tangential motion while tolerating minor mismatches. Such conditions have been modelled in contexts such as viscoelastic layers \cite{ARGATOV20113201}, interfacial sliding \cite{MIKHASEV2024104158}, and (radially-layered) piezoelectric composites \cite{Dhua16022025,DASTJERDI2025104236}. Such conditions also play a wider role when considering composite construction or applying coatings, for example imperfect bonding in composites \cite{BIGONI19983239}, weakly compressible interphases \cite{Mishuris2004ImperfectTC}, and nonhomogeneous anisotropic layers \cite{Mishuris2004ModeIII}. Similar principles apply to heat conduction in composites with interfacial resistance \cite{Lipton1996}.}

{Despite this extensive literature, the torsional vibration of a vertically coupled hollow cylinder has not been addressed. This paper fills that gap by analysing a two-section hollow cylinder under torsional loading at its upper face, with the base fixed. Three coupling types - ideal, soft, and rigid - are considered. The primary motive for this is to develop a complete qualitative and quantitative understanding of the mechanics and trade-offs associated with each type of contact in joined hollow cylinders. The study also investigates the impact of coupling type on the resonance frequencies of the cylinder. Finally, it demonstrates how a weak interface can approximate internal damage within a single-layered hollow cylinder. These results have direct implications for turbine design, automotive, aerospace and biomedical engineering, structural health monitoring, and non-destructive testing.}

The paper is organised as follows. First the problem of coupled hollow cylinders is formally stated in Sect.~\ref{Sect:ProbOverview}. The governing equations are provided, and the various contact conditions outlined. Next, in Sect.~\ref{Sect:Deriving} we introduce the modified finite Hankel transform, which will be used to solve the system and obtain the displacement. The transforms definition and properties are provided, the transformed equations stated, and the Green's function for the problem derived. The end result is an expression for the displacement in terms of the Green's function, with some unknown terms. In Sect.~\ref{Sect:ObtainingDisp}, an explicit expression for the displacement in each of the differing contact scenarios is obtained. This is used to provide numerical results in Sect.~\ref{Sect:Results}. The impact of the interface type is investigated, as well as its dependence on the cylinder geometry, material composition, and the {vibration} frequency. {As part of this the cylinders' resonance frequencies are determined, and the effect of the coupling condition on these frequencies is examined.} This model is then used in Sect.~\ref{Sect:DamageApprox} to investigate approximating damage within a single cylinder, for example the presence of a ring crack, using a weak interfacial layer. It is shown that the shear modulus of this layer needed for an effective approximation can be easily obtained from a linear relation between the shear modulus and the arithmetic mean of the displacements (observed/ ideal). Finally, concluding remarks are provided in Sect.~\ref{Sect:Conclusion}. 

\section{Problem statement
}\label{Sect:ProbOverview}

\subsection{The coupled cylinder}

\begin{figure}[t]
	\centering
	\includegraphics[width=0.45\textwidth]{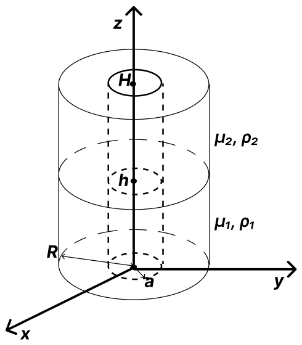}
	\caption{Geometry and coordinate system of two-layered hollow cylinder.
	}
	\label{CylinderFig}
\end{figure}

We consider the case of a coupled hollow cylinder, whose distinct upper and lower sections are connected by some interfacial layer, as depicted in Fig.~\ref{CylinderFig}. The system is considered in cylindrical coordinates $(r,\phi,z)$. {The cylinder is assumed to be {vibrating} at some frequency $\omega$.} The two sections both have inner radius $a$ and outer radius $R$, giving the problem radial coordinate $a<r<R$. They are axisymmetric about the angle $-\pi < \phi < \pi$. The lower cylinder covers height $0<z<h$, with shear modulus $\mu_1$ and density $\rho_1$. The upper cylinder covers height $h+h_0<z<H$ and has shear modulus $\mu_2$ and density $\rho_2$, where $h_0$ is the height of the interface between the two cylindrical sections. Throughout this paper we will denote the lower section using indices $k=1$, and upper cylinder $k=2$. A coupling condition describes the contact between the two sections, with the three considered cases (ideal, soft and rigid contact) being outlined in Sect.~\ref{Sect:Coupling}.

It is assumed that the cylindrical surfaces are free from load
\begin{equation}
	\label{free_surfaces}
	\left.\tau_{r\phi}^{(k)} \right|_{r=a,R} = \mu_k \left. \left( \frac{\partial u_k}{\partial r} - \frac{1}{r} u_k \right) \right|_{r=a,R} = 0, \quad k=1,2,
\end{equation}
where $\tau_{r\phi}^{(k)} (r,z)$, $u_k (r,z)$, are the tangential stress and displacement of each section $k =1,2$.

The bottom boundary of the cylinder is fixed in place, while an axisymmetric tangential torsion loading is applied to the upper face
\begin{equation} \label{upper_loading}
	u_1 (r,0) = 0, \quad \left.\tau_{z\phi}^{(2)}\right|_{z=H} = p(r) e^{i\omega t},
\end{equation}
where $\omega$ is a frequency of the cylinders {vibration}, $p(r)$ is a prescribed function, and $t$ is time. Note that this temporal term is only introduced to ensure we describe a steady-state system, and the resulting displacement will be independent of time. 

The equation of motion describing the displacement $u_k$ in each section $k=1,2$ {of the {vibrating} cylinder} is given by \cite{PopovGreens}
\begin{equation}
	\label{motion_eqn}
	\frac{1}{r} \frac{\partial}{\partial r}\left(r \frac{\partial u_k}{\partial r}\right) - \frac{1}{r^2} u_k + \frac{\partial^2 u_k}{\partial z^2} + \frac{\omega^2}{c_{k}^2} u_k = 0 , \quad k=1,2,
\end{equation}
where $c_{k} = \sqrt{\mu_k \rho_k^{-1}}$, $k=1,2$, are shear wave speeds. 

Note that in the case with $\omega=0$ expression \eqref{motion_eqn} reduces to the Laplacian, and the model can be altered to describe heat flux in a coupled cylinder, such as those used when drilling wells for geothermal energy extraction. 

\subsection{Coupling of the two layers}\label{Sect:Coupling}

We consider three types of coupling conditions between the cylinder sections: ideal contact, soft (weak) contact, and rigid (hard/ stiff) contact. The boundary formulation for this coupling needs to be defined in order to construct the general solution of the boundary value problem \eqref{free_surfaces}-\eqref{motion_eqn}. We therefore outline the three distinct couplings below.

\subsubsection{Ideal contact}

The case of ideal contact assumes perfect coupling between the two cylinder sections, such that the displacement $u_k$ and tangential torsion $\tau_{z\phi}^{(k)}$ of the two sections $k=1,2$ at the boundary are identical. The ideal contact conditions between the sections can therefore be described in terms of the jump $[\![ . ]\!]$ as
\begin{equation} \label{ideal_contact}
	\left[\![ u_k (r,h) ]\!\right]  = u_2(r,h+0) - u_1 (r, h - 0) = 0,
\end{equation}
$$
[\![ \tau_{z\phi}^{(k)}(r,h) \, \!]\!] = \tau_{z\phi}^{(2)}(r,h+0) - \tau_{z\phi}^{(1)} (r,h-0)  = \mu_2 \left.\frac{\partial u_2}{\partial z}\right|_{z=h+0} - \mu_1 \left. \frac{\partial u_1}{\partial z}\right|_{z=h-0} = 0.
$$

\subsubsection{Soft contact}

The case of soft (weak) contact assumes that there exists some thin interface between the upper and lower sections, which is significantly weaker than the cylindrical layers themselves. The interface is therefore defined by its density $\rho_0$, its height $h_0 \ll H$ (the interface is thin), and its shear modulus $\mu_0 \ll \mu_k$, $k=1,2$ (it is constructed of a soft material). An example of such an interface would be the use of a weak adhesive which deforms but maintains its integrity through the process, or a layer added to provide energy/ thermal dissipation. 

The contact conditions for such a thin ``soft" interface are provided in  \cite{Mishuris2004ImperfectTC}. It is assumed that there is continuity of the stress function over the interface, while the jump of displacements is supposed proportional to the stress
\begin{equation}
	\label{soft_contact}
	\left[\![ u_k (r,h) ]\!\right]  = K_1 \left.\frac{\partial u_2}{\partial z}\right|_{z=h+0} , \quad \![\![ \tau_{z\phi}^{(k)}(r,h)\, \!]\!] = 0,
\end{equation}
where $K_1 = \mu_2 \mu_0^{-1} h_0$.

\subsubsection{Rigid contact}

The case of rigid (hard/ stiff) contact assumes that the two cylindrical sections are coupled via some significantly stronger interface layer{, which restricts lateral motion but allows for force transmission}. The interface is again defined by its density $\rho_0$, its height $h_0 \ll H$ (the interface is thin), and its shear modulus $\mu_0 \gg \mu_k$, $k=1,2$ (the interface is strong/ rigid). Examples of rigid contact would include the bonding of a metal layer between rubber sections, as used in some wheel designs.

The contact conditions for such a rigid interface are provided in \cite{Mishuris2004ImperfectTC}. The displacements across the interface are assumed to be continuous, such that the displacement jump is zero.
\begin{equation}\label{rigid_continuity}
	\![\![ u_k (r,h) \, \!]\!] = 0.
\end{equation}
Meanwhile, the equation of motion for the interfacial layer takes the form
\begin{equation}
	\label{rigid_interface}
	\frac{1}{r}\frac{d}{dr}\left( r \frac{dv(r)}{dr} \right) - \frac{1}{r^2}v(r) + \frac{1}{\mu_0 h_0}\left( \mu_2 \left. \frac{\partial u_2}{\partial z}\right|_{z=h+0} - \mu_1 \left.\frac{\partial u_1}{\partial z}\right|_{z=h-0} \right) + \frac{\omega^2}{c_0^2}v(r) = 0,
\end{equation}
where $v(r)$ is displacement of the interface, and $c_0=\sqrt{\mu_0 \rho_0^{-1}}$.

It is also assumed that the cylindrical surfaces of the interface are free from stress
\begin{equation}
	\label{rigid_free_stress}
	\left.\tau_{r\phi}^{(0)} \right|_{r=a,R} = \mu_0 \left.\left[ \frac{dv(r)}{dr} - \frac{1}{r}v(r) \right] \right|_{r=a,R} = 0,
\end{equation}
where $\tau_{r\phi}^{(0)}$ is the tangential stress of the rigid interface.

\section{The deriving of a solution in the transform’s domain}\label{Sect:Deriving}

With the problem geometry and interface conditions outlined, we now seek to obtain a unique expression for the displacements $u_{k}$, $k=1,2$, for each case. This will be achieved by first applying the finite Hankel transform, whose definition and properties are outlined in Sect.~\ref{Sect:HankelTransform}. Applying this transform eliminates the variable $r$, with the reduction of the transformed system to a one dimensional boundary value problem provided in Sect.~\ref{Sect:Reduced}. This transformed problem can then be solved in terms of a 1D Green's function, which is obtained in Sect.~\ref{Sect:Greens}. 

\subsection{The finite Hankel transform}\label{Sect:HankelTransform}

In order to reduce the problem \eqref{free_surfaces}-\eqref{motion_eqn} to a one dimensional boundary value problem, we utilize a variant of the finite Hankel transform, defined as (see e.g. \cite{FiniteHankel,Martynenko1986}):
\begin{equation}
	\label{integral_transform}
	u_{kn}(z) = \int_a^R u_k (r,z) K\left(\lambda_n, r\right) r \, dr , \quad k=1,2, \quad n\in\mathbb{N}_0,
\end{equation}
with the kernel
$$
K\left(\lambda_n , r\right) = J_2 (\lambda_n a) Y_1 (\lambda_n r) - Y_2 (\lambda_n a) J_1 (\lambda_n r),
$$
where $J_{1,2} (.)$, $Y_{1,2} (.)$ are Bessel functions of the first and second kind respectively (see e.g. \cite{Korenev_Bessel}), and $\lambda_n$ are nonnegative roots of the transcendental equation \cite{FiniteHankel}
$$
J_2 (\lambda a) Y_2 (\lambda R) - Y_2 (\lambda a) J_2 (\lambda R) = 0.
$$
It is necessary to underline that due to the special boundary conditions \eqref{free_surfaces} $\lambda_0 = 0$ is also an eigenvalue with the corresponding eigenfunction $y_0(r) = r$ and norm \cite{FiniteHankel}
$$
\| y_0 (r) \|^2 = \int_a^R r^2 \cdot r \, dr = \frac{R^4 - a^4}{4}.
$$ 

When applied to the governing equation \eqref{motion_eqn}, the finite Hankel transform reduces it to a 1D problem due to the following result
\begin{equation}\label{transform_derivatives}
	\int_a^R  \left[\frac{\partial}{\partial r}\left(r \frac{\partial u_k}{\partial r}\right)  - \frac{1}{r}  u_k \right] 
	K\left(\lambda_n, r\right)  \, dr  =  \lambda_n^2  u_{kn}(z), 
\end{equation}
which follows from taking integration by parts and utilizing the properties of Bessel functions (see e.g. \cite{Korenev_Bessel}).

Once the solution of the transformed problem is obtained, the original system is recovered utilizing the inverse finite Hankel transform as
\begin{equation}
	\label{inverse_transform}
	u_k (r,z) = \frac{4r}{R^4 - a^4} u_{k0} (z) + \sum_{n=1}^\infty u_{kn}(z) \frac{K(\lambda_n,r)}{\| K(\lambda_n,r)\|^2},
\end{equation}
where
$$
\| K(\lambda_n,r)\|^2 = \int_a^R K^2(\lambda_n,r) r\, dr = \frac{R^2}{2}K^2(\lambda_n,r) - \frac{2}{\pi^2 \lambda_n^2}, \quad n\in\mathbb{N}.
$$

\subsection{The transformed problem}\label{Sect:Reduced}

With the finite Hankel transform introduced, we apply it to the governing equations and contact conditions for the system. This will yield a transformed problem that we can more easily solve.

\subsubsection{The governing equations}

Applying the finite Hankel transform \eqref{integral_transform} to the governing system of equations \eqref{free_surfaces}-\eqref{motion_eqn}, noting \eqref{transform_derivatives}, yields
\begin{equation}
	\label{transformed_problem}
	\left\{
	\begin{array}{l r}
		u_{kn}^{\prime\prime} (z) - \gamma_{kn}^2 u_{kn}(z) = 0 , \quad & 0<z<H, \quad z\neq h, \\[2mm]
		u_{1n}(0) = 0, \quad  u_{2n}^\prime (H) = \mu_2^{-1}p_n, &
	\end{array}
	\right.
\end{equation}
where $n\in\mathbb{N}_0$ and
$$
\gamma_{kn}^2 = \lambda_n^2 - \frac{\omega^2}{c_k^2}, \quad p_n = \int_a^R p(r) K(\lambda_n,r) r\, dr .
$$

While this provides the transformed system for all $n\in\mathbb{N}_0$, in the following analysis it is useful to take advantage of the known $\lambda_0 = 0$ and consider the case $n=0$ separately. Evaluating \eqref{transformed_problem} for $n=0$ gives
\begin{equation} \label{transformed_prob0}
	\left\{
	\begin{array}{l r}
		u_{k0}^{\prime\prime}(z) + \beta_k^2 u_{k0}(z) = 0 , \quad & 0<z<H, \quad z\neq h, \\[2mm]
		u_{10}(0) = 0 , \quad u_{20}^{\prime}(H) = \mu_2^{-1}p_0, &
	\end{array}
	\right.
\end{equation}
where
$$
\beta_k^2 = \frac{\omega^2}{c_k^2}, \quad p_0 = \int_a^R p(r) r^2 \, dr,
$$
where we have used that $K(0,r)=y_0(r) = r$.

This is part of the transformed system that we seek to solve. The remainder will come from the form of the contact condition between the two cylinders. 

\subsubsection{The coupling conditions}

Alongside transforming the governing equations, we must also transform the coupling conditions for the three types of contact being considered. Let us handle each of these separately.\\

\noindent {\bf Ideal contact:} In transform’s domain \eqref{integral_transform} the conditions \eqref{ideal_contact} take the form
\begin{equation}\label{ideal_transformed}
	\![\![ u_{kn}(h) \, \!]\!] = 0, \quad \![\![ u_{kn}^\prime (h) \, \!]\!] = \frac{\mu_1 - \mu_2}{\mu_1} u_{2n}^\prime (h+0) , \quad n\in\mathbb{N}_0.
\end{equation}
\smallskip

\noindent {\bf Soft contact:} The coupling conditions \eqref{soft_contact} in transform’s domain \eqref{integral_transform} have the following form
\begin{equation}
	\label{soft_transformed}
	\![\![ u_{kn}(h) \, \!]\!] = K_1 u_{2n}^\prime (h+0), \quad \![\![ u_{kn}^\prime (h) \, \!]\!] = \frac{ \mu_1 - \mu_2}{\mu_1} u_{2n}^\prime (h+0), \quad n\in\mathbb{N}_0.
\end{equation}
\smallskip

\noindent {\bf Rigid contact:} Applying the finite Hankel transform \eqref{integral_transform} to the coupling conditions for rigid contact \eqref{rigid_continuity}-\eqref{rigid_free_stress} yields the transformed conditions
\begin{equation}
	\label{rigid_interface_trans}
	\![\![ u_{kn}(h)\, \!]\!] = 0, \quad n\in\mathbb{N}_0, \quad
	\![\![ u^\prime_{kn}(h) \, \!]\!] =  \frac{\mu_1 - \mu_2}{\mu_1} u_{2n}^\prime (h+0) + \frac{\mu_0 h_0 \gamma_{0n}^2}{\mu_1} v_n, \quad n\in\mathbb{N}_0,
\end{equation}
where $\beta_0^2 = \frac{\omega^2}{c_0^2}$, and $\gamma_{0n}^2 = \lambda_n^2 - \beta_0^2$. 
In order to maintain coupling of the displacement between the two cylinders and the rigid layer, it is supposed that
$$
u_{1n}(h-0) = u_{2n}(h+0) = v_n.
$$
Consequently, again treating the case $n=0$ separately, the second condition in \eqref{rigid_interface_trans} can be rewritten in the form
\begin{equation} \label{rigid_interface_trans_better}
	\begin{aligned}
		\![\![ u^\prime_{k0}(h) \, \!]\!] &= \frac{\mu_1 - \mu_2}{\mu_1} u^\prime_{20}(h+0) - K_2 \beta_0^2 u_{20}(h+0), \\[2mm]
		\![\![ u^\prime_{kn}(h) \, \!]\!] &= \frac{\mu_1 - \mu_2}{\mu_1} u^\prime_{2n}(h+0) + K_2 \gamma_{0n}^2 u_{2n}(h+0), \quad n\in\mathbb{N},
	\end{aligned}
\end{equation}
where $K_2 = \mu_0 h_0 \mu_1^{-1}$.\\

With these conditions obtained, we are now in a position to begin solving the transformed problem. This is achieved by considering the Green's function associated for the boundary value problem.

\subsection{A Green's function representation of the transformed displacement}\label{Sect:Greens}

\subsubsection{The Green's function}

The Green's function $G_{kn}(z,\eta)$, $k=1,2$, $n\in\mathbb{N}_0$, in the transformed domain is the solution of (see e.g. \cite{PopovGreens})
\[
\left\{
\begin{array}{l r}
	\frac{\partial^2 G_{kn}(z,\eta)}{\partial z^2} - \gamma_{kn}^2 G_{kn}(z,\eta) = \delta(z-\eta), \quad & 0<z,\eta < H, \quad z,\eta \neq h,\\[2mm]
	G_{1n}(0,\eta) = 0 , \quad & 0<\eta <H,\\[2mm]
	\left.\frac{\partial G_{2n}(z,\eta)}{\partial z}\right|_{z=H} = 0 , \quad & 0<\eta <H.
\end{array}
\right.
\]
This is solved using standard techniques. 
The resulting fundamental basic system of solutions and Green’s function \cite{PopovGreens} are then given by 
\begin{equation} \label{Greens_Func}
	\Psi_{0n}(z) = \frac{\cosh\left(\gamma_{2n}H - \gamma_{kn}z\right)}{\cosh\left(\gamma_{2n}H\right)} , \quad \Psi_{1n}(z) = \frac{\sinh\left(\gamma_{2n}z\right)}{\gamma_{2n} \cosh\left( \gamma_{2n} H\right)} ,
\end{equation}
$$
G_{kn}(z,\eta) = -\frac{1}{\gamma_{kn} \cosh\left(\gamma_{2n}H\right)} 
\left\{ 
\begin{array}{l r}
	\cosh\left(\gamma_{2n} H - \gamma_{kn}\eta \right) \sinh\left(\gamma_{kn} z\right) , \quad & z<\eta, \\[2mm]
	\cosh\left(\gamma_{2n} H - \gamma_{kn}z\right) \sinh\left(\gamma_{kn}\eta \right) , \quad & z>\eta .
\end{array}	
\right.
$$

While the above holds for all $n\in\mathbb{N}_0$, it will again be useful to consider the case $n=0$ separately. Evaluating  \eqref{transformed_prob0} for $n=0$ the fundamental basic system of solutions and Green’s function for the problem can be written as
\begin{equation} \label{Greens_Func0}
	\Psi_{00}(z) = \frac{\cos\left(\beta_2 H - \beta_k z\right)}{\cos(\beta_2 H)} , \quad \Psi_{10}(z) = \frac{\sin(\beta_2 z)}{\beta_2 \cos\left(\beta_2 H\right)},
\end{equation}
$$
G_{k0}(z,\eta) = -\frac{1}{\beta_k \cos\left(\beta_2 H\right)}
\left\{
\begin{array}{l r}
	\cos\left(\beta_2 H - \beta_k \eta\right) \sin(\beta_k z) , \quad & z<\eta,\\[2mm]
	\cos\left(\beta_2 H - \beta_k z\right) \sin(\beta_k \eta) , \quad & z>\eta.
\end{array}	
\right.
$$

Expressions \eqref{Greens_Func}, \eqref{Greens_Func0} provide the Green's function for the problem, however they do not account for the coupling condition at the interface. 

\subsubsection{Representing the transformed displacement in terms of the Green's function}

To incorporate the coupling conditions at the interface $z=h$, we seek to represent the transformed displacement $u_{kn}$ in terms of the Green's function. To determine the proper representation of the transformed displacement, let us consider some of the properties of the Green's function. 

It has previously been established (see e.g. \cite{PopovGreens}) that the Green’s function $G_{kn}(z,\eta)$ has the following discontinuous properties during the transition from the point $z=h-0$ to the point $z=h+0$: \\
\begin{itemize}
	\item the Green's function is continuous
	$$
	\![\![ G_{kn}(z,h)\, \!]\!] = 0 ;
	$$
	\item its derivative has the jump equal to $1$
	$$
	\![\![ \frac{\partial G_{kn}(z,h)}{\partial z} \, \!]\!] = 1 ;
	$$
	\item its derivative with respect to variable $\eta$ has jump equal to $-1$
	$$
	\![\![ \frac{\partial G_{kn}(z,h)}{\partial \eta} \, \!]\!] = -1 ;
	$$ 
	\item the jump of the Green's functions second order mixed derivative is zero
	$$
	\![\![ \frac{\partial^2 G_{kn}(z,h)}{\partial z \partial \eta} \, \!]\!] = 0.
	$$  
\end{itemize}

This inspires searching for the problem’s solution as a superposition of its continuous and discontinuous parts. Following this approach, the solution of \eqref{transformed_problem}, \eqref{transformed_prob0} are derived in terms of the jump of the transformed displacements’ $\![\![ u_{kn}(h)\, \!]\!]$ and their derivatives $\![\![ u_{kn}^\prime (h)\, \!]\!]$ as
\begin{equation}
	\label{displacement}
	u_{kn} = \frac{p_n \sinh(\gamma_{2n}z)}{\mu_2 \gamma_{2n} \cosh(\gamma_{2n}H)} + \, \![\![ u_{kn}^\prime (h) \, \!]\!] \left. G_{kn}(z,\eta)\right|_{\eta=h} - \, \![\![ u_{kn}(h)\, \!]\!] \left.\frac{\partial G_{kn}(z,\eta)}{\partial \eta}\right|_{\eta=h} ,
\end{equation}
where $n\in\mathbb{N}$, while for the case $n=0$
\begin{equation}
	\label{displacement0}
	u_{k0} = \frac{p_0 \sin(\beta_{2}z)}{\mu_2 \beta_{2} \cos(\beta_{2}H)} + \, \![\![ u_{k0}^\prime (h)\, \!]\!] \left. G_{k0}(z,\eta)\right|_{\eta=h} - \, \![\![ u_{k0}(h)\, \!]\!] \left.\frac{\partial G_{k0}(z,\eta)}{\partial \eta}\right|_{\eta=h} .
\end{equation}
The form of the Green's function is already known from \eqref{Greens_Func}, \eqref{Greens_Func0}. Meanwhile, the remaining terms $\![\![ u_{kn}(h) \, \!]\!]$ and $\![\![ u_{kn}^\prime (h) \,\!]\!]$ need to be computed from the specific contact condition.

\section{Obtaining an explicit expression for the displacement}\label{Sect:ObtainingDisp}

With the form of the displacement specified \eqref{displacement}-\eqref{displacement0}, all that remains is to obtain the unknown transformed displacement jumps $\left[\left[ u_{kn}(h) \right]\right]$ and $\left[\left[ u_{kn}^\prime (h)\right]\right]$, $k=1,2$. We obtain these from the coupling conditions associated with the specified contact condition \eqref{ideal_transformed}-\eqref{rigid_interface_trans_better}, before applying the inverse finite Hankel transform \eqref{inverse_transform} to obtain the displacement. We consider each of these separately, with the derivation and result for ideal contact being given in Sect.~\ref{Sect:Ideal}, soft contact in Sect.~\ref{Sect:Soft}, and rigid contact in Sect.~\ref{Sect:Rigid}.

\subsection{The displacement for ideal coupling of the cylinders}\label{Sect:Ideal}

Combining the formulas for the displacements \eqref{displacement}-\eqref{displacement0} with the ideal contact conditions \eqref{ideal_transformed}, we obtain the system to be solved for the case of ideal contact
\begin{equation}\label{ideal_pre_displacement}
	u_{kn}(z) = \frac{p_n \sinh\left(\gamma_{2n}z\right)}{\mu_2 \gamma_{2n} \cosh\left(\gamma_{2n} H\right)} - \frac{\mu_2 - \mu_1}{\mu_1} G_{kn}(z,h)u_{2n}^\prime (h+0), \quad n\in\mathbb{N},
\end{equation}
\begin{equation}
	\label{ideal_pre_displacement0}
	u_{k0}(z) = \frac{p_0 \sin\left(\beta_{2}z\right)}{\mu_2 \beta_{2} \cos\left(\beta_{2} H\right)} - \frac{\mu_2 - \mu_1}{\mu_1} G_{k0}(z,h)u_{20}^\prime (h+0),
\end{equation}
In order to find the unknown values $u_{2n}^\prime(h+0)$ and $u_{20}^\prime (h+0)$ the expressions \eqref{ideal_pre_displacement}-\eqref{ideal_pre_displacement0} are differentiated with respect to variable $z$, and the substitutions $k=2$ and $z=h+0$ are made. The resulting linear algebraic equations are then solved to obtain
\begin{equation}
	u_{20}^\prime (h+0) = \frac{p_0 \cos\left(\beta_2 h\right)}{\mu_2 \Delta_{20}^I}, \quad u_{2n}^\prime(h+0) = \frac{p_n \cosh\left(\gamma_{2n}h\right)}{\mu_2 \Delta_{2n}^I}, \quad n\in\mathbb{N},
\end{equation}
where
$$
\Delta_{20}^I = \cos\left(\beta_2 H\right) - \frac{\mu_2 - \mu_1}{\mu_1} \sin\left(\beta_2 h\right)\sin\left[\beta_2 \left(H-h\right)\right] ,
$$
$$
\Delta_{2n}^I = \cosh\left(\gamma_{2n}H\right) + \frac{\mu_2 - \mu_1}{\mu_1} \sinh\left(\gamma_{2n}h\right)\sinh\left[\gamma_{2n}\left(H-h\right)\right], \quad n\in\mathbb{N}.
$$
Inserting into  \eqref{ideal_pre_displacement}-\eqref{ideal_pre_displacement0}, we obtain the explicit expressions for the transformed displacements
\begin{equation}\label{ideal_displacement_trans}
	u_{kn}(z) = \frac{p_n \sinh\left(\gamma_{2n}z\right)}{\mu_2 \gamma_{2n} \cosh\left(\gamma_{2n} H\right)} - \frac{\mu_2 - \mu_1}{\mu_1}  \frac{p_n \cosh\left(\gamma_{2n}h\right)}{\mu_2\Delta_{2n}^I} G_{kn}(z,h), \quad n\in\mathbb{N},
\end{equation}
\begin{equation}
	\label{ideal_displacement0_trans}
	u_{k0}(z) = \frac{p_0 \sin\left(\beta_{2}z\right)}{\mu_2 \beta_{2} \cos\left(\beta_{2} H\right)} - \frac{\mu_2 - \mu_1}{\mu_1} \frac{p_0 \cos\left(\beta_2 h\right)}{\mu_2 \Delta_{20}^I} G_{k0}(z,h),
\end{equation}
Finally, applying the inverse finite Hankel transform \eqref{inverse_transform} to this expression yields 
\begin{equation}
	\label{ideal_displacement}
	\begin{aligned}
		u_k (r,z) =& \frac{4rp_0}{\mu_2 (R^4 - a^4)}\left[ \frac{\sin\left(\beta_2 z\right)}{\beta_2 \cos\left(\beta_2 H\right)} - \frac{\mu_2 - \mu_1}{\mu_1} \frac{\cos\left(\beta_2 h\right)}{\Delta_{20}^I} G_{k0}(z,h)\right]  \\
		+& \frac{1}{\mu_2}\sum_{n=1}^\infty p_n \left[ \frac{\sinh\left(\gamma_{2n}z\right)}{\gamma_{2n}\cosh\left(\gamma_{2n}H\right)} - \frac{\mu_2 - \mu_1}{\mu_1} \frac{\cosh\left(\gamma_{2n}h\right)}{\Delta_{2n}^I} G_{kn}(z,h) \right] \frac{K(\lambda_n,r)}{\|K(\lambda_n,r)\|^2}, \quad k=1,2.
	\end{aligned}
\end{equation}
This is the explicit expression for the displacements in the case of ideal contact. This can also be used to obtain the unique expression for the stress $\tau_{z\phi}^{(k)}$, $k=1,2$.


\subsection{The displacement for soft coupling between the cylinders}\label{Sect:Soft}

Utilizing the transformed conditions for soft contact \eqref{soft_transformed}, the expressions for the displacements \eqref{displacement}-\eqref{displacement0} can be written in the form
\begin{equation}
	\label{soft_pre_displacement}
	u_{kn}(z) = \frac{p_n \sinh\left(\gamma_{2n}z\right)}{\mu_2 \gamma_{2n} \cosh\left(\gamma_{2n}H\right)} - \left[ K_1 \left. \frac{\partial G_{kn}(z,\eta)}{\partial \eta}\right|_{\eta=h} + \frac{\mu_2 - \mu_1}{\mu_1} G_{kn}(z,h) \right] u_{2n}^{\prime} (h+0), \quad n\in\mathbb{N},
\end{equation} 
\begin{equation}
	\label{soft_pre_displacement0}
	u_{k0}(z) = \frac{p_0 \sin\left(\beta_{2}z\right)}{\mu_2 \beta_{2} \cos\left(\beta_{2}H\right)} - \left[ K_1 \left. \frac{\partial G_{k0}(z,\eta)}{\partial \eta}\right|_{\eta=h} + \frac{\mu_2 - \mu_1}{\mu_1} G_{k0}(z,h) \right] u_{20}^{\prime} (h+0),
\end{equation} 

The unknown values $u_{2n}^\prime (h+0)$ and $u_{20}^\prime (h+0)$ are obtained using an almost identical procedure to that for the case of ideal contact. Namely, expressions \eqref{soft_pre_displacement}-\eqref{soft_pre_displacement0} are differentiated with respect to variable $z$, and the substitutions $k=2$ and $z=h+0$ are made. The resulting linear algebraic equations for $u_{2n}^\prime (h+0)$ are solved to yield
\begin{equation}
	\label{soft_disder_trans}
	u_{20}^\prime (h+0) =  \frac{p_0\cos\left(\beta_2 h\right)}{\mu_2 \Delta_{20}^S} , \quad u_{2n}^\prime (h+0) = \frac{p_n \cosh\left(\gamma_{2n}h\right)}{\mu_2 \Delta_{2n}^S} ,
\end{equation}
where
$$
\Delta_{20}^S = \cos\left(\beta_2 H\right) + \sin\left[\beta_2 \left(H-h\right)\right]\left[ K_1 \beta_2 \cos\left(\beta_2 h\right) + \frac{\mu_2 - \mu_1}{\mu_1} \sin\left(\beta_2 h\right) \right],
$$
$$
\Delta_{2n}^S = \cosh\left(\gamma_{2n} H\right) + \sinh\left[\gamma_{2n} \left(H-h\right)\right]\left[ K_1 \gamma_{2n} \cosh\left(\gamma_{2n} h\right) + \frac{\mu_2 - \mu_1}{\mu_1} \sinh\left(\gamma_{2n} h\right) \right].
$$

Inserting into \eqref{soft_pre_displacement}-\eqref{soft_pre_displacement0}, the explicit form of displacements in the transform domain is obtained 
\begin{equation}
	\label{soft_displacement_trans}
	u_{kn}(z) = \frac{p_n \sinh\left(\gamma_{2n}z\right)}{\mu_2 \gamma_{2n} \cosh\left(\gamma_{2n}H\right)} -  \left[ K_1 \left.\frac{\partial G_{kn}(z,\eta)}{\partial \eta}\right|_{\eta=h} + \frac{\mu_2 - \mu_1}{\mu_1} G_{kn}(z,h)\right] \frac{p_n\cosh\left(\gamma_{2n}h\right)}{\mu_2 \Delta_{2n}^S},
\end{equation}
\begin{equation}
	\label{soft_displacement_trans0}
	u_{k0}(z) = \frac{p_0 \sin\left(\beta_2 z\right)}{\mu_2 \beta_2 \cos\left(\beta_2 H\right)} -  \left[ K_1 \left.\frac{\partial G_{k0}(z,\eta)}{\partial \eta}\right|_{\eta=h} + \frac{\mu_2 - \mu_1}{\mu_1} G_{k0}(z,h)\right] \frac{p_0\cos\left(\beta_2 h\right)}{\mu_2 \Delta_{20}^S}.
\end{equation}
It can be seen that when constant $K_1=0$ the transformed displacements \eqref{soft_displacement_trans}-\eqref{soft_displacement_trans0} coincide with those for the ideal contact subcase \eqref{ideal_displacement_trans}-\eqref{ideal_displacement0_trans}.

Applying the inverse finite Hankel transform \eqref{inverse_transform} to the expressions for the transformed displacements \eqref{soft_displacement_trans}-\eqref{soft_displacement_trans0} yields an explicit expression for the displacements in the case of soft contact
{\small 
	\begin{equation}
		\label{soft_displacement}
		\begin{aligned}
			&u_k (r,z) = \frac{4r p_0}{\mu_2 (R^4 - a^4)}\left[ \frac{\sin(\beta_2 z)}{\beta_2 \cos(\beta_2 H)} - \left( K_1 \left.\frac{\partial G_{k0}(z,\eta)}{\partial \eta}\right|_{\eta=h} + \frac{\mu_2-\mu_1}{\mu_1} G_{k0}(z,h) \right) \frac{\cos(\beta_2 h)}{\Delta_{20}^S} \right] \\
			+& \frac{1}{\mu_2}\sum_{n=1}^\infty p_n \left[ \frac{\sinh(\gamma_{2n}z)}{\gamma_{2n}\cosh(\gamma_{2n}H)} + \left( K_1 \left.\frac{\partial G_{kn}(z,\eta)}{\partial \eta}\right|_{\eta=h} + \frac{\mu_2 -\mu_1}{\mu_1} G_{kn}(z,h)\right) \frac{\cosh(\gamma_{2n}h)}{\Delta_{2n}^S} \right] \frac{K(\lambda_n, r)}{\| K(\lambda_n , r)\|^2} ,
		\end{aligned}
\end{equation}}
with $k=1,2$. This can again be utilized to obtain a unique expression for the stress $\tau_{z\phi}^{(k)}$, $k=1,2$.

\subsection{The displacement for rigid coupling between the cylinders}\label{Sect:Rigid}

Combining the transformed rigid contact conditions \eqref{rigid_interface_trans_better} with the assumed form of the displacement \eqref{displacement}-\eqref{displacement0}, the discontinuous problem for rigid contact is given by
\begin{equation}
	\label{rigid_disc_prob}
	u_{kn}(z) = \frac{p_n \sinh(\gamma_{2n}z)}{\mu_2 \gamma_{2n}\cosh(\gamma_{2n}H)} - \left[ \frac{\mu_2-\mu_1}{\mu_1} u_{2n}^\prime (h+0) - K_2 \gamma_{0n}^2 u_{2n}(h+0) \right] G_{kn}(z,h) ,
\end{equation}
\begin{equation}
	\label{rigid_disc_prob0}
	u_{k0}(z) = \frac{p_0 \sin(\beta_{2}z)}{\mu_2 \beta_{2}\cos(\beta_{2}H)} - \left[ \frac{\mu_2-\mu_1}{\mu_1} u_{20}^\prime (h+0) + K_2 \beta_{0}^2 u_{20}(h+0) \right] G_{k0}(z,h) ,
\end{equation}
We now seek the unknown values $u_{2n}(h+0)$ and $u_{2n}^\prime (h+0)$ using same approach as in the two previous subcases. First the expressions \eqref{rigid_disc_prob}-\eqref{rigid_disc_prob0} are differentiated, followed by making the substitutions $k=2$ and $z=h+0$. The resulting system of linear algebraic equations for the unknown values $u_{2n}(h+0)$ and $u_{2n}^\prime(h+0)$ are then solved for the cases $n=0$ and $n\in\mathbb{N}$, yielding
\begin{equation}
	\begin{array}{l l}
		u_{20}(h+0) = \displaystyle\frac{p_0 \sin(\beta_2 h)}{\mu_1 \Delta_{20}^{R}} , \quad & u_{20}^\prime (h+0) = \displaystyle\frac{p_0}{\mu_2\Delta_{20}^{R}} \left[ \beta_2 \cos(\beta_2 h) - K_2 \beta_0^2 \sin(\beta_2 h)\right], \\[4mm]
		u_{2n}(h+0) = \displaystyle\frac{p_n \sinh(\gamma_{2n} h)}{\mu_1 \Delta_{2n}^{R}} , \quad & u_{2n}^\prime (h+0) = \displaystyle\frac{p_n}{\mu_2\Delta_{2n}^{R}} \left[ \gamma_{2n} \cosh(\gamma_{2n} h) + K_2 \gamma_{0n}^2 \sinh(\gamma_{2n} h)\right], 
	\end{array}
\end{equation}
where $n\in\mathbb{N}$ and
$$
\Delta_{20}^{R} = \beta_2 \cos(\beta_2 H) - \left[ \frac{\mu_2 - \mu_1}{\mu_1} \beta_2 \sin\left[\beta_2(H-h)\right] + K_2 \beta_0^2 \cos\left[\beta_2 (H-h)\right]\right]\sin(\beta_2 h),
$$
$$
\Delta_{2n}^{R} = \gamma_{2n} \cosh(\gamma_{2n} H) + \left[ \frac{\mu_2 - \mu_1}{\mu_1} \gamma_{2n} \sinh\left[\gamma_{2n}(H-h)\right] + K_2 \gamma_{0n}^2 \cosh\left[\gamma_{2n} (H-h)\right]\right]\sinh(\gamma_{2n} h).
$$
Inserting into \eqref{rigid_disc_prob}-\eqref{rigid_disc_prob0}, the transformed displacements are obtained
\begin{equation}
	\label{rigid_displacement_trans}
	u_{kn}(z) = \frac{p_n \sinh(\gamma_{2n}z)}{\mu_2 \gamma_{2n}\cosh(\gamma_{2n}H)} - \frac{p_n}{\mu_2 \Delta_{2n}^{R}}\left[ \frac{\mu_2 - \mu_1}{\mu_1} \gamma_{2n} \cosh(\gamma_{2n}h) - K_2 \gamma_{0n}^2 \sinh(\gamma_{2n}h)\right] G_{kn}(z,h),
\end{equation}
\begin{equation}
	\label{rigid_displacement_trans0}
	u_{k0}(z) = \frac{p_0 \sin(\beta_{2}z)}{\mu_2 \beta_{2}\cos(\beta_{2}H)} - \frac{p_0}{\mu_2 \Delta_{20}^{R}}\left[ \frac{\mu_2 - \mu_1}{\mu_1} \beta_{2} \cos(\beta_{2}h) - K_2 \beta_{0}^2 \sin(\beta_{2}h)\right] G_{k0}(z,h).
\end{equation}
Note again that when the constant $K_2=0$ the derived solutions \eqref{rigid_displacement_trans}-\eqref{rigid_displacement_trans0} coincide with those for the subcase of ideal contact \eqref{ideal_displacement_trans}-\eqref{ideal_displacement0_trans}. 

Applying the inverse finite Hankel transform \eqref{inverse_transform} to the transformed displacements \eqref{rigid_displacement_trans}-\eqref{rigid_displacement_trans0} yields an explicit expression for the displacements
{\small \begin{equation}
		\label{rigid_displacement}
		\begin{aligned}
			&u_k (r,z) = \frac{4rp_0}{\mu_2 (R^4 - a^4)} \left[ \frac{\sin(\beta_2 z)}{\beta_2 \cos(\beta_2 H)} - \frac{G_{k0}(z,h)}{\Delta_{20}^{R}}\left( \frac{\mu_2-\mu_1}{\mu_1} \beta_2 \cos(\beta_2 h) - K_2 \beta_0^2 \sin(\beta_2 h)\right) \right] \\[2mm]
			&+ \frac{1}{\mu_2}\sum_{n=1}^\infty p_n\left[ \frac{\sinh(\gamma_{2n}z)}{\gamma_{2n}\cosh(\gamma_{2n}H)} - \frac{G_{kn}(z,h)}{\Delta_{2n}^{R}}\left( \frac{\mu_2 - \mu_1}{\mu_1} \gamma_{2n} \cosh(\gamma_{2n}h) - K_2 \gamma_{0n}^2 \sinh(\gamma_{2n}h) \right)\right] \frac{K(\lambda_n,r)}{\|K(\lambda_n,r)\|^2} ,
		\end{aligned}
\end{equation}}
where $k=1,2$. This can again be utilized to obtain a unique expression for the stress $\tau_{z\phi}^{(k)}$, $k=1,2$.

\section{Numerical results for the displacement and stress}\label{Sect:Results}

{Having determined unique expressions for the displacements obtained for all three cases: \eqref{ideal_displacement}, \eqref{soft_displacement}, and \eqref{rigid_displacement} for ideal, soft, and rigid contact, respectively, the tangential stress then follows from the standard stress-displacement relations. In the following sections, unless stated otherwise we consider a cuprum-aluminium pair, the material properties being provided in Table.~\ref{Table:Material}. The interphase layer is taken such that $h_0=0.01$ [m]; $\mu_{\text{soft}}=(\mu_1+\mu_2)/200$, $\mu_{\text{rigid}}=100(\mu_1+\mu_2)/2$, and $\rho_0=(\rho_1+\rho_2)/2$. The upper surface loading is $p(r)=(1-r^2)$ (steady-state), so 
	\begin{equation}\label{results_loading} \left. \tau_{z\phi}^{(2)}\right|_{z=H} = (1-r^2)e^{i\omega t}.
	\end{equation}
	We set $\omega=1$ [Hz] in Sections~\ref{Sect:ResultsInterface}--\ref{Sect:ResultsParameters} and explore frequency effects in Section~\ref{Sect:Resultsvibration}. Simulations conducted for other cylinder geometries and material compositions were found to match the general trends reported in this section, and thus are omitted for the sake of brevity.}

	\begin{table}[t]
		\centering
		\begin{tabular}{|c||c|c|c|c|}
			\hline
			& Height & Density $\rho$ & Shear modulus $\mu$ & Wave-speed $c$  \\
			& [m] & [kg/m$^3$] & [Pa] & [m/s] \\
			\hline \hline 
			Lower section & $0.05$ & $8900$ & $40.70 \times 10^9$ & 2250 \\
			(Cuprum) & & & & \\
			\hline
			Upper section &  & $2700$ & $25.00\times 10^9$ & 3110 \\
			(Aluminium) & & & & \\
			\hline 	
			Soft interface & $0.01$ & $5800$ & $3.285 \times 10^9$ & 2680 \\
			\hline
			Rigid interface & $0.01$ & $5800$ & $3285 \times 10^9$ & 2680 \\
			\hline
		\end{tabular}
		\caption{Material properties and geometry of the cylinder used in simulations for Sect.~\ref{Sect:ResultsInterface}. The height of the upper cylinder is $0.05$ [m] in the case of ideal contact, and $0.04$ [m] in the case of a soft or rigid interface, to maintain a fixed total height $H=0.1$ [m]. The hollow cylinders' inner radius is $a=0.25$ [m], while the outer radius is $R=0.5$ [m].}
		\label{Table:Material}
	\end{table}
	
	In Sect.~\ref{Sect:ResultsInterface}, we consider the impact that the type of interface (ideal, soft or rigid) has on the resulting cylinder displacements and stresses. Then, in Sect.~\ref{Sect:ResultsParameters} we examine the impact of varying the shear modulus of the interface layer in the case of soft and rigid contact. Finally, the varying of {vibration} rates is considered in Sect.~\ref{Sect:Resultsvibration}, {with investigations into both its impact on the radial distribution of the displacement and stress, and the determining of the resonance frequencies of the cylinder}.
	
	\subsection{Impact of the interface on the system behaviour}\label{Sect:ResultsInterface}
	
\begin{figure}[tp!]
		\centering
		\includegraphics[width=0.45\textwidth]{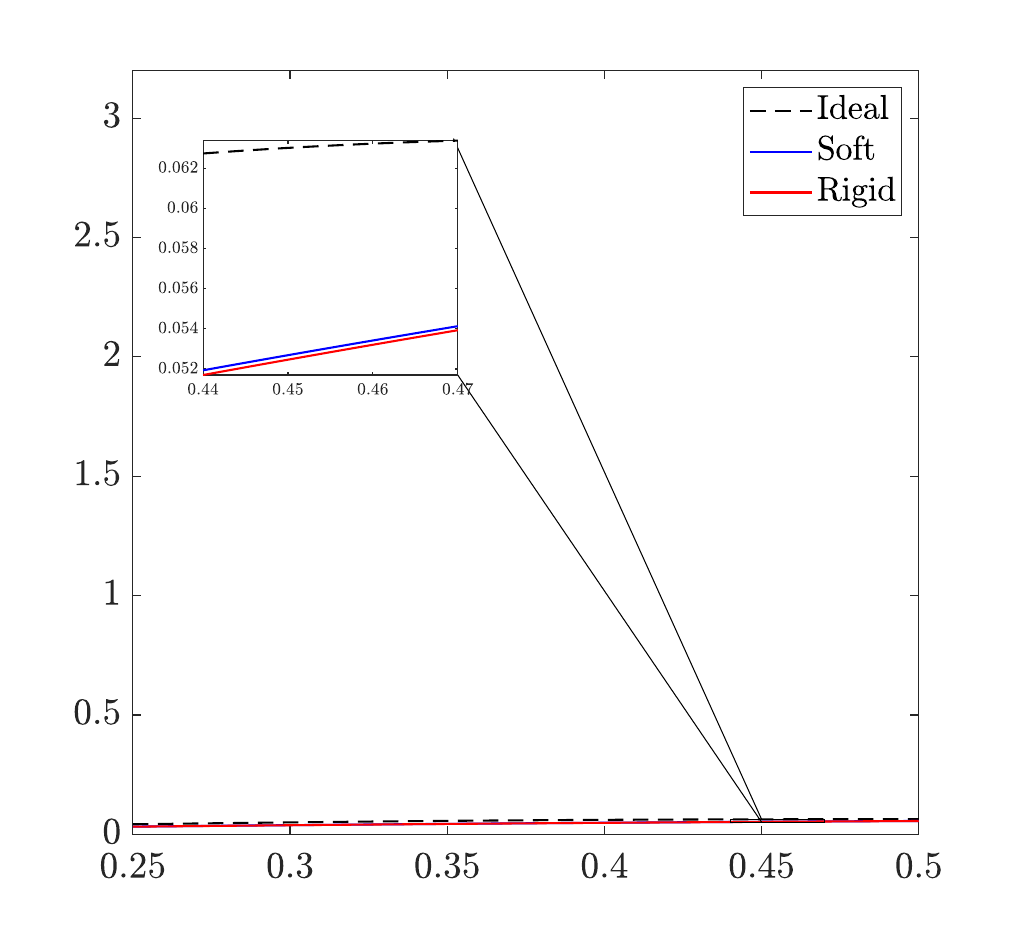}
		\put(-107,155) {$z/H=0.25$}
		\put(-85,0) {$r$}
		\put(-210, 80) {$\mu_2 u(r,z)$}
		\put(-195,159) {{\bf (a)}}
		\hspace{8mm}
		\includegraphics[width=0.45\textwidth]{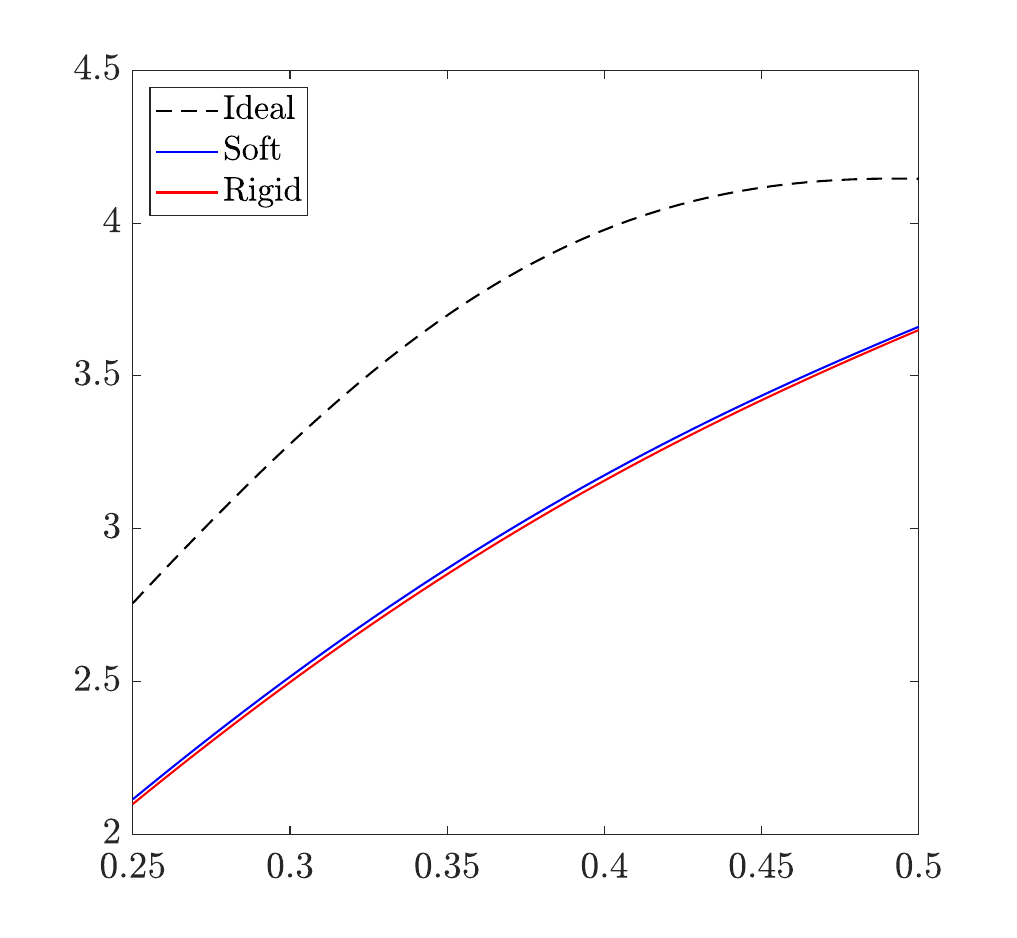}
		\put(-107,155) {$z/H=0.25$}
		\put(-85,0) {$r$}
		\put(-212, 80) {$\tau_{z\phi}(r,z)$}
		\put(-195,159) {{\bf (b)}}
		
		\vspace{4mm}
		
		\includegraphics[width=0.45\textwidth]{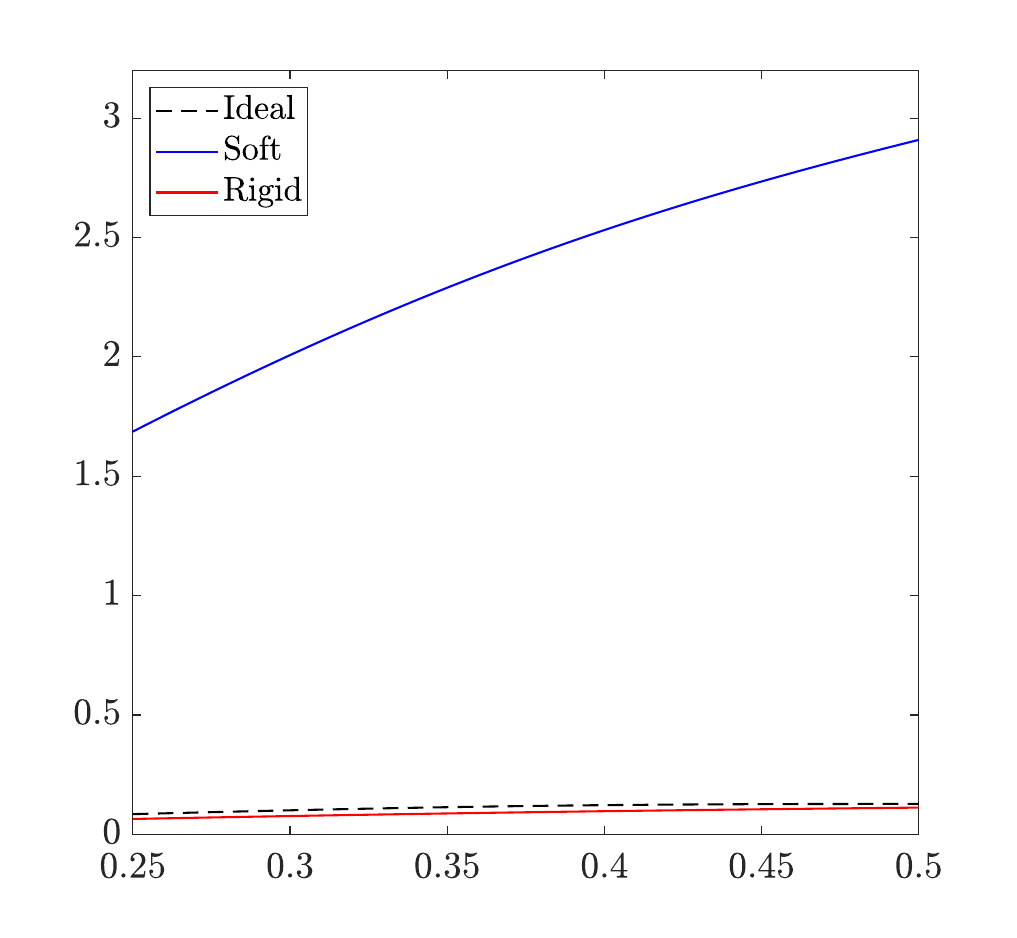}
		\put(-107,155) {$z/H=0.50$}
		\put(-85,0) {$r$}
		\put(-210, 80) {$\mu_2 u(r,z)$}
		\put(-195,159) {{\bf (c)}}
		\hspace{8mm}
		\includegraphics[width=0.45\textwidth]{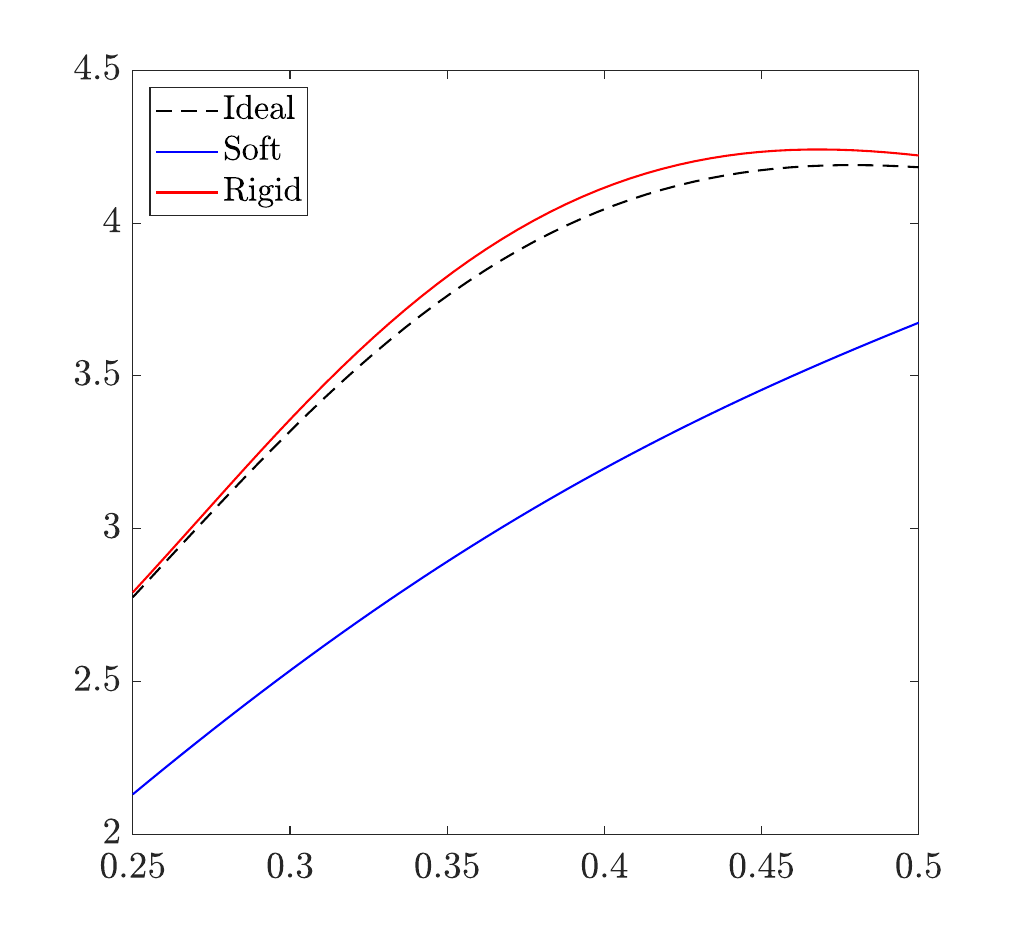}
		\put(-107,155) {$z/H=0.50$}
		\put(-85,0) {$r$}
		\put(-212, 80) {$\tau_{z\phi}(r,z)$}
		\put(-195,159) {{\bf (d)}}
		
		\vspace{4mm}
		
		\includegraphics[width=0.45\textwidth]{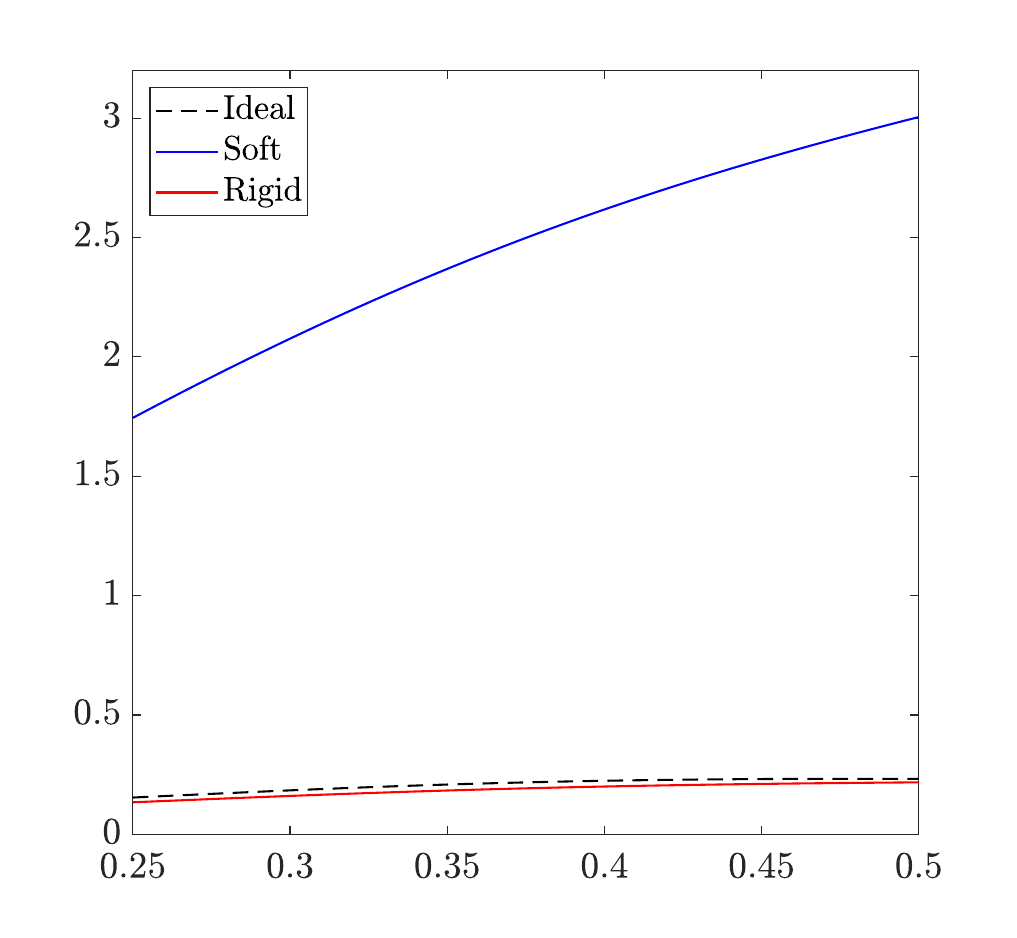}
		\put(-107,155) {$z/H=0.75$}
		\put(-85,0) {$r$}
		\put(-210, 80) {$\mu_2 u(r,z)$}
		\put(-195,159) {{\bf (e)}}
		\hspace{8mm}
		\includegraphics[width=0.45\textwidth]{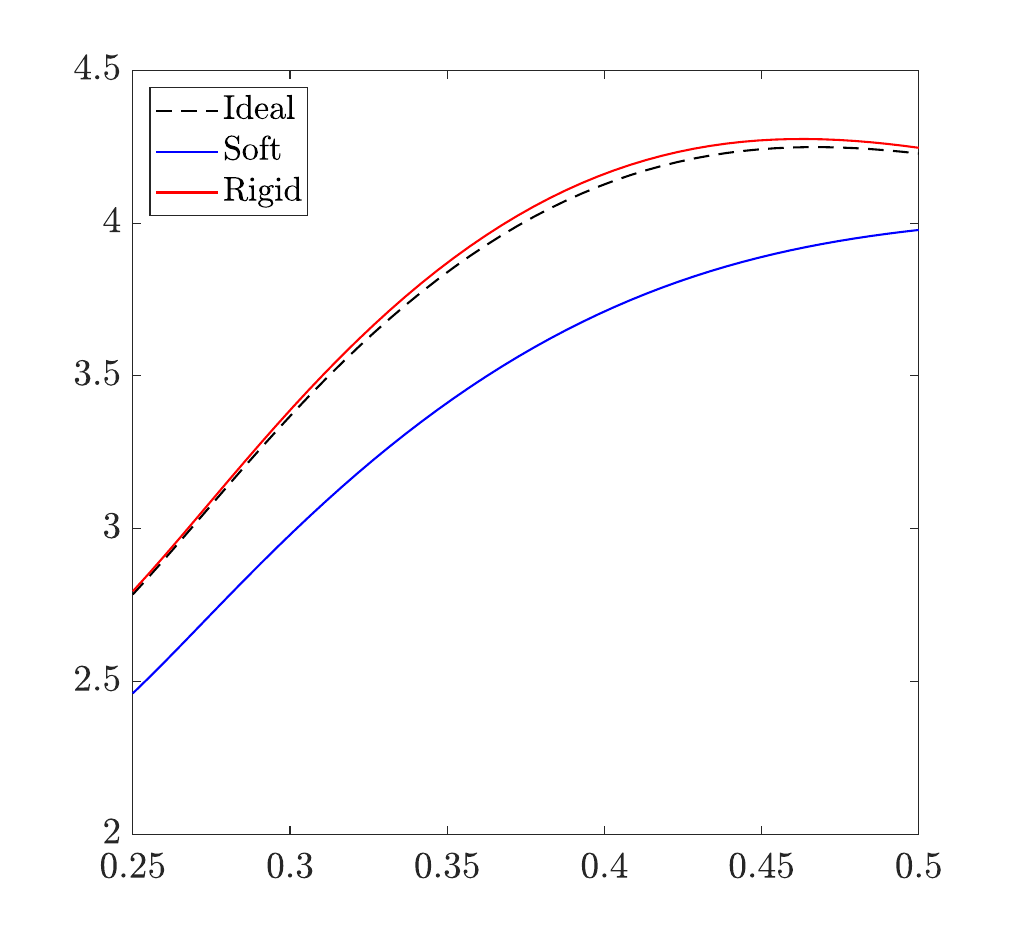}
		\put(-107,155) {$z/H=0.75$}
		\put(-85,0) {$r$}
		\put(-212, 80) {$\tau_{z\phi}(r,z)$}
		\put(-195,159) {{\bf (f)}}
		\caption{The change of (a), (c), (e) the scaled displacement and (b), (d), (f) the stress for various fixed heights in a coupled cuprum-aluminium cylinder with an ideal, soft or rigid interface between the sections. We show the fixed heights: (a), (b) the middle of the lower (cuprum) cylinder $z/H=0.25$, (c), (d) the (start of the) interface $z/H=0.5$, (e), (f) (near) the middle of the upper (aluminium) cylinder $z/H=0.75$.}
		\label{Fig_Layers_Mid}
	\end{figure}
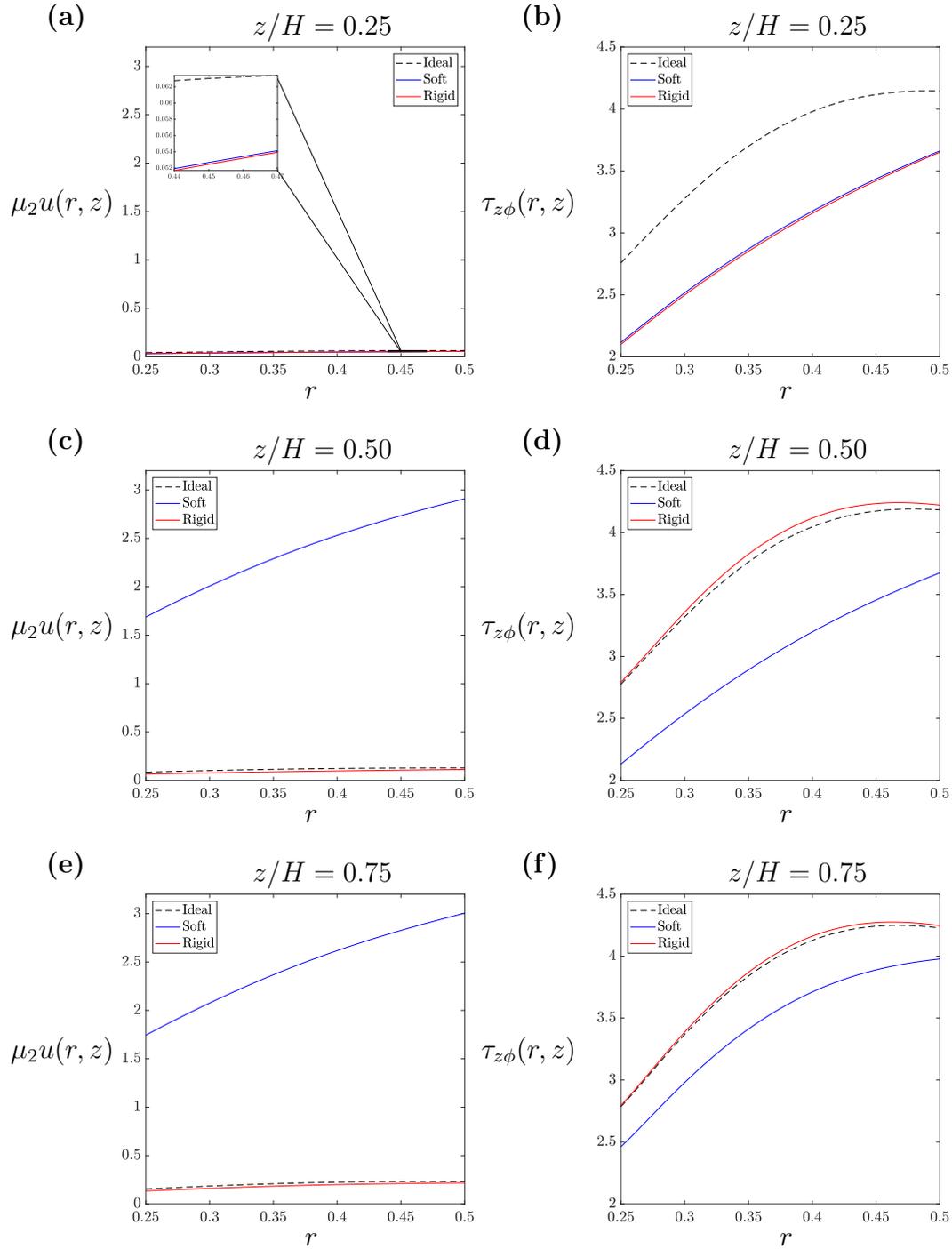

	{For the cuprum-aluminium cylinder, whose geometry and material properties are provide in Table.~\ref{Table:Material}, we compare the scaled displacement $\mu_2 u(r,z)$ and tangential stress $\tau_{z\varphi}(r,z)$ at $z/H=0.25,0.5,0.75$. Results are provided in Fig.~\ref{Fig_Layers_Mid}.}
	
	{In the lower cylinder ($z/H=0.25$, Fig.~\ref{Fig_Layers_Mid}a,b), soft and rigid contacts yield nearly identical displacements and stresses, whereas ideal contact produces a slightly different stress profile (larger magnitude and less linear over $r$). At the interface ($z/H=0.5$, Fig.~\ref{Fig_Layers_Mid}c,d), soft contact yields a significantly larger displacement but smaller tangential stress than the other two models. The result for the ideal and rigid contact conditions are very close to one another, although there is a small difference for both the displacement and stress. Finally, in the upper cylinder ($z/H=0.75$, Fig.~\ref{Fig_Layers_Mid}e,f), trends mirror those at the interface, with soft contact still producing larger displacements, while the ideal and rigid cases remain similar. The tangential stresses are closer in the upper section, due to the loading occurring at the top of the cylinder.} 
	
	{We can now consider the  dependence of the stress and displacement on the radial position $r$ seen in Fig.~\ref{Fig_Layers_Mid}. It can be seen that the displacements are nearly linear at all vertical positions in the cylinder. Meanwhile, the stresses show near-linearity in the lower cylinder for soft/rigid contact, but not ideal. They remain near-linear for soft contact at the interface, but not for ideal/rigid contact. The tangential stress becomes nonlinear above the interface in all cases, and are no longer monotonic over $r$ for the ideal/rigid coupling.}
	
	
	%
	%
	%
	
	Investigations by the authors have shown that almost-linear behaviour dominates in the case where the ratio of cylinder outer radius to height $R/H < 5$. Conversely, when taking $R/H\geq 5$ a clearly nonlinear pattern emerges for the displacement and stress. {This motivates further investigation into the impact of this ratio and the form of interfacial coupling.}

	\subsection{Impact of varying the cylinder geometry and composition}\label{Sect:ResultsParameters}
	
	\begin{table}[t]
		\centering
		\begin{tabular}{l  c}
			& {\bf Relative difference between ideal and rigid contact} \\[2mm]
			{\bf Displacement} &
			\begin{tabular}{|l || c | c | c |}
				\hline \hline
				& $10.0$ & $50.0$ & $100.0$
				\\ \hline \hline                                   
				0.5 & 0.002160 & 0.004938 & 0.005884 \\
				\hline
				2.5 & 0.065772 & 0.163634 & 0.201234 \\
				\hline
				5.0 & 0.062962 & 0.180386 & 0.237041 \\
				\hline \hline
			\end{tabular}
			\put(-232, 0) {$R/H$}
			\put(-115,42) {$\mu_0 / \mu_1$} \\
			&\\
			{\bf Stress} &
			\begin{tabular}{|l || c | c | c |}
				\hline \hline
				& $10.0$ & $50.0$ & $100.0$
				\\ \hline \hline                                   
				0.5 & 0.004433 & 0.010135 & 0.012076 \\
				\hline
				2.5 &  0.016629 & 0.040485 & 0.049558\\
				\hline
				5.0 &  0.005923 & 0.014839 & 0.018881\\
				\hline \hline
			\end{tabular}
			\put(-232, 0) {$R/H$}
			\put(-115,42) {$\mu_0 / \mu_1$}\\
			&\\[-2mm]
		\end{tabular}
		\caption{Relative difference between the cases of ideal and rigid contact for the (a) displacement $u(r,z)$, (b) tangential stress $\tau_{z\phi}(r,z)$, at the start of the cylinders' interface $z=h$ and the radial centre of the hollow cylinder $r=(R+a)/2$.  Three values of soft interface shear modulus $\mu_0$ are considered, while the upper and lower cylinder shear moduli $\mu_1=10^9$ [Pa], $\mu_2=2\times 10^9$ [Pa], are kept fixed between simulations. Results are shown for varying cylinder width-height ratio $R/H$, where $H=0.1$ [m] is the (fixed) total height of the cylinder, while outer radius $R$ is varied, and the inner radius is taken as $a=R/2$ [m]. }
		\label{RigidDiff}
	\end{table}
	
	
	{We examine the impact of the aspect ratio $R/H$ on the solution by keeping fixed height $H=0.1$ [m] and varying the cylinder outer radius $R$ to obtain $R/H=0.5,2.5,5.0$, taking inner radius $a=R/2$ [m] in all cases. Similarly, we investigate the effect of interfacial stiffness by fixing $\mu_1=10^9$ [Pa], $\mu_2=2\times 10^9$ [Pa], and varying the ratio of the interface shear modulus $\mu_0/\mu_1$ (for soft or rigid contact).  The remaining parameters $\rho_0,\rho_1,\rho_2$ and $h_0$ follow Table.~\ref{Table:Material}. We report the relative difference of the displacement and tangential stress versus the ideal-contact baseline at the radial centre $r=(R+a)/2$ and start of the interface $z=h$.}
	
	
	
	The relative difference between the case of ideal contact and that with a rigid interface layer is provided in Table.~\ref{RigidDiff}. It can again be seen that there is a close correspondence between the these cases (as previously shown graphically in Fig.~\ref{Fig_Layers_Mid}), with the final relative difference between them depending on both the interface ratio $\mu_0/\mu_1$ and the width-height ratio $R/H$. The difference in the tangential stress between ideal contact and the rigid interface is close to $1$\% for the cases $R/H = 0.5, 5.0$, for all values of the shear modulus ratio $\mu_0 / \mu_1$. Interestingly, there is a value of $R/H$ for which the relative difference in the tangential stress between rigid and ideal contact is maximised at the interface $z=h$, with the relative differences for $R/H=2.5$ being consistently higher than those for $R/H=0.5$ or $R/H=5$, reaching almost $5$\% when $\mu_0 / \mu_1 = 100$. The relative differences in the displacement meanwhile behaves monotonically, increasing with both $R/H$ and $\mu_0 / \mu_1$. While the difference is small for the case of a long, thin cylinder ($R/H=0.5$) - never exceeding $0.6$\% - the case of a short, fat cylinder ($R/H=5.0$) sees a relative difference of the displacement as high as $24$\%.

	\begin{table}[t]
		\centering
		\begin{tabular}{l  c}
			& {\bf Relative difference between ideal and soft contact} \\[2mm]
			{\bf Displacement} &
			\begin{tabular}{|l || c | c | c |}
				\hline \hline
				& 0.1 & 0.05 & 0.01
				\\ \hline \hline                                   
				0.5 &  0.651907 & 1.30039 & 6.47525 \\
				\hline
				2.5 &  2.88752 & 5.49900 & 25.0973 \\
				\hline
				5.0 &  5.78273 & 10.8609 & 46.6705 \\
				\hline \hline
			\end{tabular}
			\put(-232, 0) {$R/H$}
			\put(-115,42) {$\mu_0 / \mu_1$} \\
			&\\[-2mm]
			{\bf Stress} &
			\begin{tabular}{|l || c | c | c |}
				\hline \hline
				& 0.1 & 0.05 & 0.01
				\\ \hline \hline                                   
				0.5 &  0.011179 & 0.015338 & 0.021838 \\
				\hline
				2.5 &  0.117186 & 0.166791 & 0.252894 \\
				\hline
				5.0 &  0.109589 & 0.166652 & 0.290020 \\
				\hline \hline
			\end{tabular}
			\put(-232, 0) {$R/H$}
			\put(-115,42) {$\mu_0 / \mu_1$}\\
			&\\[-2mm]
		\end{tabular}
		\caption{Relative difference between the cases of ideal and soft contact for the (a) displacement $u(r,z)$, (b) tangential stress $\tau_{z\phi}(r,z)$, at the start of the cylinders' interface $z=h$ and the radial centre of the hollow cylinder $r=(R+a)/2$.  Three values of soft interface shear modulus $\mu_0$ are considered, while the upper and lower cylinder shear moduli $\mu_1=10^9$ [Pa], $\mu_2=2\times 10^9$ [Pa], are kept fixed between simulations. Results are shown for varying cylinder width-height ratio $R/H$, where $H=0.1$ [m] is the (fixed) total height of the cylinder, while outer radius $R$ is varied, and the inner radius is taken as $a=R/2$ [m].}
		\label{SoftDiff}
	\end{table}
	
	Next, the relative difference between the case of ideal contact and that with a soft interface layer is provided in Table.~\ref{SoftDiff}. There is again a clear difference in behaviour between the soft interface and that for ideal or rigid contact. There is a monotonic increase in the relative difference of the displacement with increasing width-height ratio $R/H$ and shear moduli ratio $\mu_0 / \mu_1$. The differences are however far larger than those for the comparison of rigid and ideal contact conditions, with the minimum relative difference of the displacement at $z=h$ being $65$\%, and the maximum exceeding $4667$\%. Note that this exceptionally high value is due to the cylinder material composition, geometry, and very low shear modulus of the soft layer - with the value in the other considered cases being far lower. Meanwhile, the relative difference of the tangential stress increases monotonically with decreasing shear moduli ratio $\mu_0 / \mu_1$, but follows a more complicated pattern with varying width-height ratio $R/H$. When $\mu_0/\mu_1 = 0.1, 0.05$ there is a maximum of the relative difference (compare $R/H=2.5$ with $R/H=0.5, 5.0$), however it behaves monotonically with $R/H$ when $\mu_0 / \mu_1 = 0.01$. Regardless, in all cases the relative difference of the tangential stress between the soft and ideal contact conditions are significantly larger than those for the rigid-ideal comparison, with the minimum again being about $1$\%, but the maximum exceeding $29$\% at the interface $z=h$.

	\subsection{Impact of varying the cylinder {vibration} rate} \label{Sect:Resultsvibration}
	
	In the proceeding subsections we kept a fixed cylinder {vibration} rate $\omega = 1$ [Hz]. Now, let us vary the {vibration} rate, and examine the impact on the stress and displacement for the three types of coupled interface.
	
	{This consists of two separate investigations. In Sect.~\ref{Sect:VibDist} we investigate the impact of the vibration rate on the displacement and stress behaviour over $r$ {away from the resonance frequencies}. Then, in Sect.~\ref{Sect:Resonance} we utilize the model to investigate the resonance {frequencies of the cylinder} under torsion loading, as well as the impact of the assumed coupling.}
	
	\subsubsection{Impact of vibration rate {away from resonance frequencies}}\label{Sect:VibDist}
	
	We again consider an cuprum-aluminium cylinder {whose geometry and material properties are as stated in Table.~\ref{Table:Material}, while the torsion loading on the upper face of the cylinder is again given by \eqref{results_loading}.} The results provided below are however general to all cylinder geometries and material compositions considered by the authors. {In this subsection it is assumed that the vibrational frequency is not close to any resonance frequency of the cylinder.}
	
	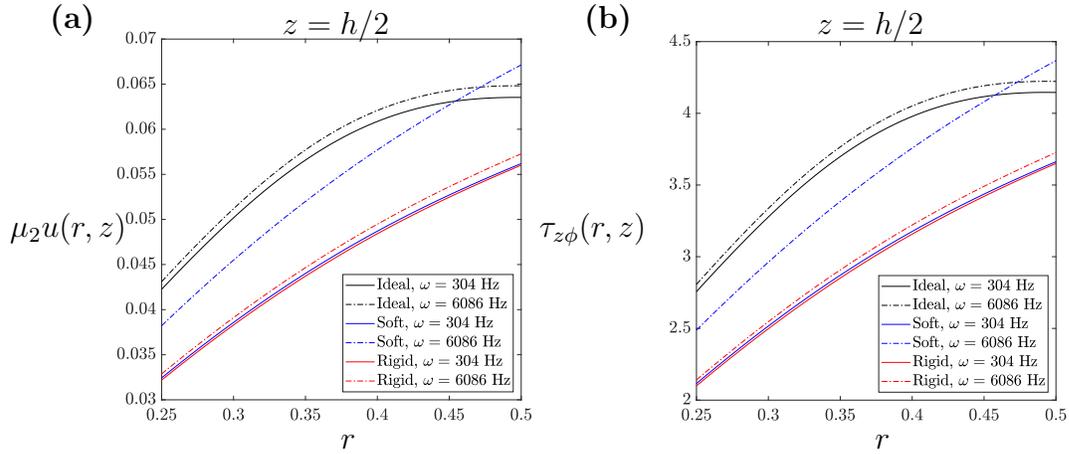
\begin{figure}[tp!]
		\centering
		\includegraphics[width=0.45\textwidth]{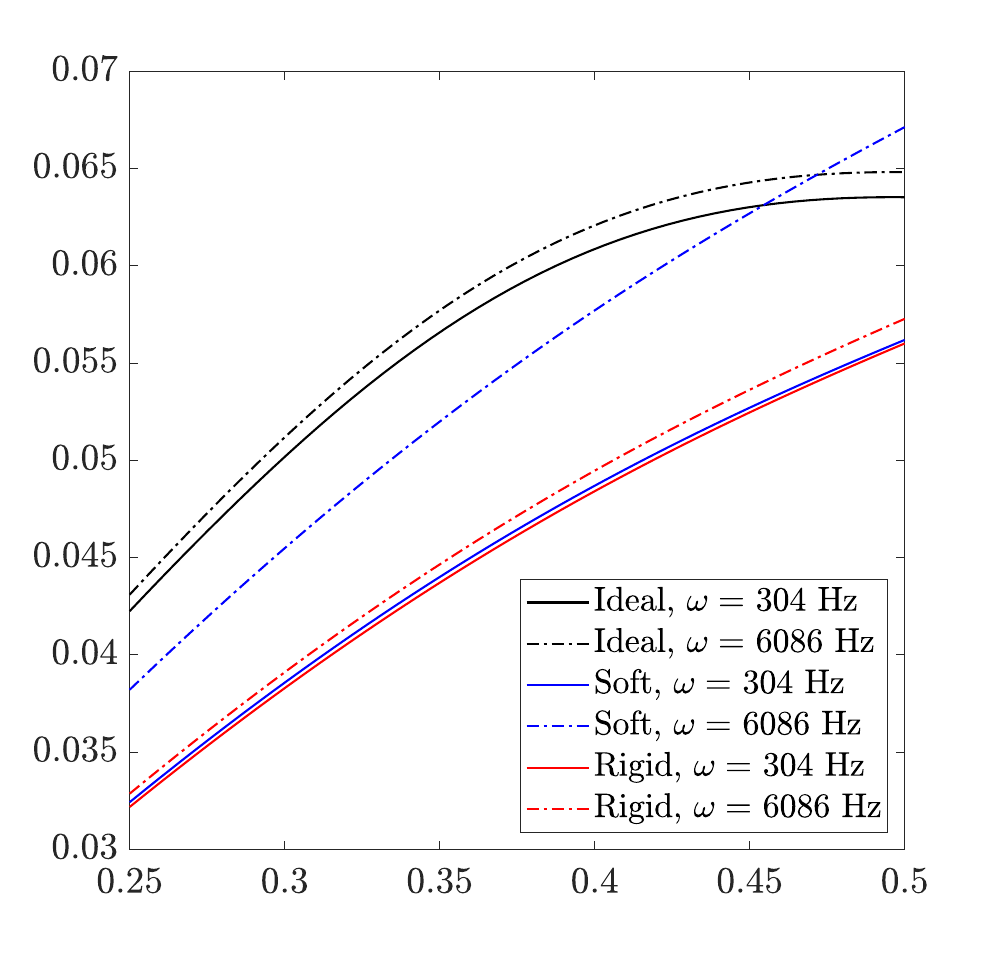}
		\put(-107,157) {$z=h/2$}
		\put(-85,0) {$r$}
		\put(-210, 80) {$\mu_2 u(r,z)$}
		\put(-195,159) {{\bf (a)}}
		\hspace{8mm}
		\includegraphics[width=0.45\textwidth]{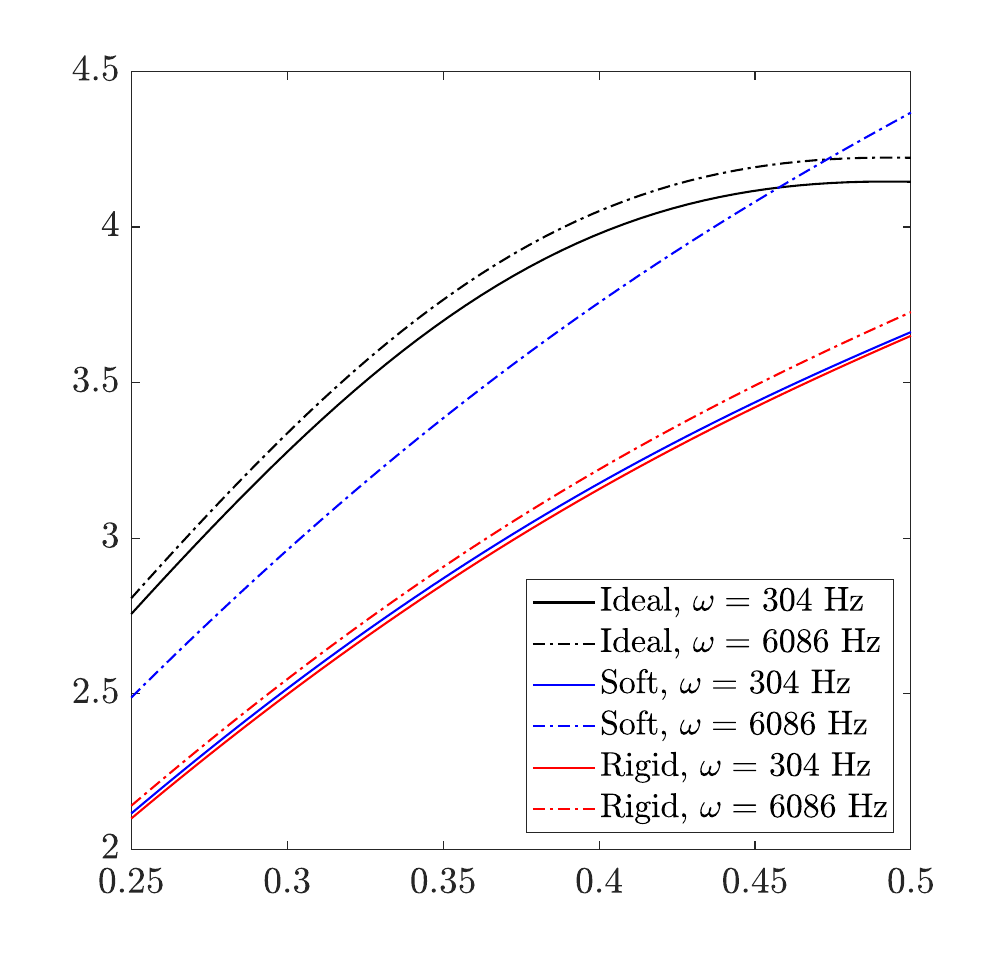}
		\put(-107,157) {$z=h/2$}
		\put(-85,0) {$r$}
		\put(-212, 80) {$\tau_{z\phi}(r,z)$}
		\put(-195,159) {{\bf (b)}}
		\caption{The change of (a) the scaled displacement and (b) the stress in a coupled cuprum-aluminium cylinder subject to differing {vibration} rates $\omega$ with an ideal, soft or rigid interface between the upper and lower cylinders. All values are taken at the fixed height $z=h/2$ (middle of the lower cylinder).}
		\label{Fig_vibration}
	\end{figure}
	
	Results for the differing interfacial layers (ideal, soft, rigid) under two differing {vibration} rates are provided in Fig.~\ref{Fig_vibration}. Note that these correspond to the same height as shown in Fig.~\ref{Fig_Layers_Mid}a,b (for $\omega = 1$ [Hz]). The {vibration} rates in Fig.~\ref{Fig_vibration} are chosen such that the natural normalisation $\tilde{\omega} = \omega H/\sqrt{\mu_2 / \rho_2}$ takes values $\tilde{\omega}=0.01$ and $\tilde{\omega}=0.2$.
	
	As would be expected, increasing the {vibration} rate leads to an outwards (positive) displacement and tangential stress. Comparing Fig.~\ref{Fig_Layers_Mid}a,b with Fig.~\ref{Fig_vibration}, it is clear that the effect is only marginal when going from $\omega=1$ [Hz] to $\omega=304$ [Hz], but becomes far more significant when increasing to $\omega = 6086$ [Hz]. The magnitude of this effect is also significantly larger in the case of a soft interface, compared to that for the ideal and rigid interfacial layer. {This effect holds for the range of frequencies considered here, but behaviour will vary when approaching resonance frequencies, as considered in the next section.} 
	
	\subsubsection{{The resonance frequencies}}\label{Sect:Resonance}
	
	{Alongside impacting the radial distribution of the stress and displacement, the vibrational frequency can also induce resonance-related effects that will greatly amplify the stress and displacement. In this section, we demonstrate that these resonance frequencies, $\omega^*$ can be obtained using the presented model, and investigate the impact of the coupling condition on the resulting resonance frequencies.}
	
	\begin{figure}[tp!]
		\centering
		\includegraphics[width=0.48\textwidth]{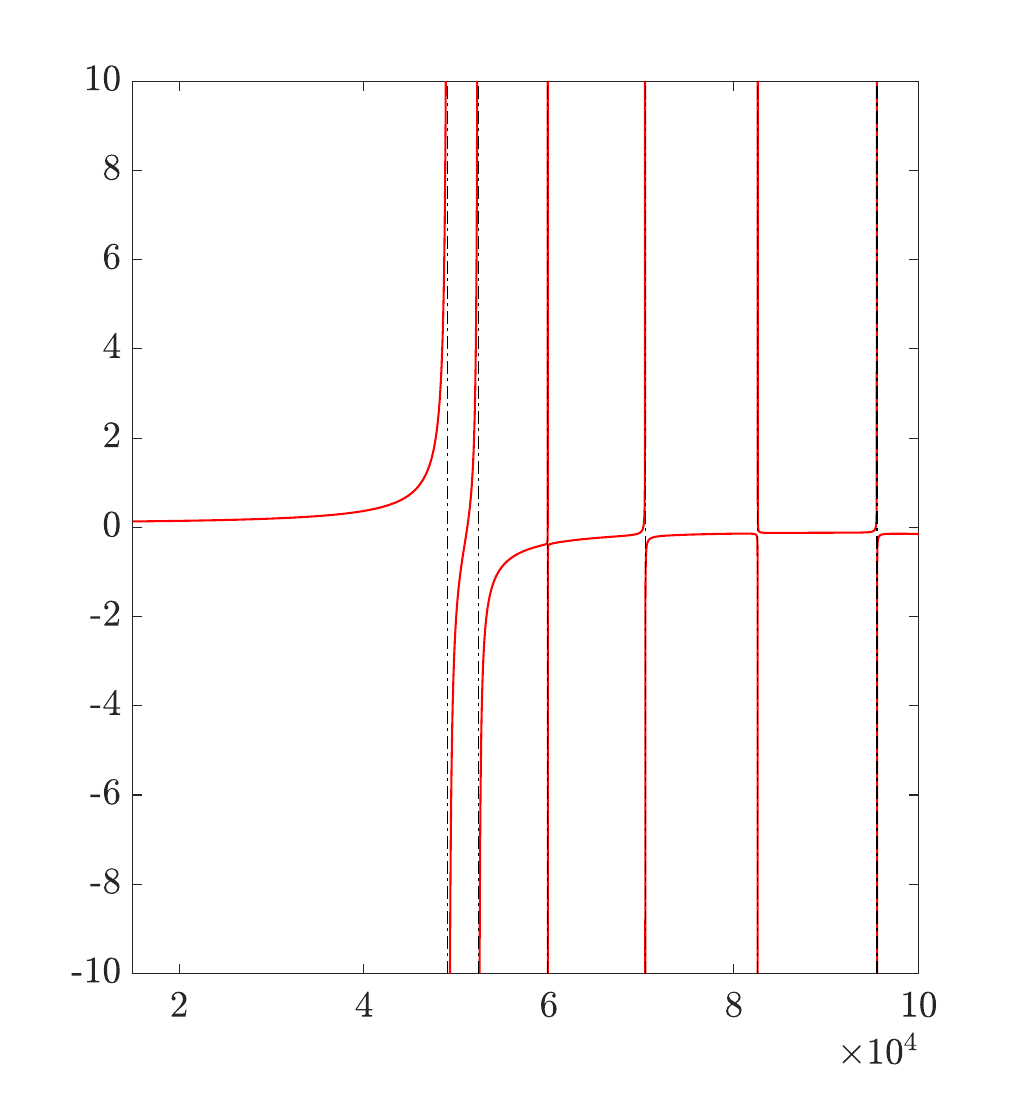}
		\put(-107,190) {Ideal}
		\put(-90,0) {${\omega}$}
		\put(-193, 80) {\rotatebox{90}{$\mu_2 u(r^*,z^*)$}}
		\put(-195,190) {{\bf (a)}}
		\hspace{4mm}
		\includegraphics[width=0.48\textwidth]{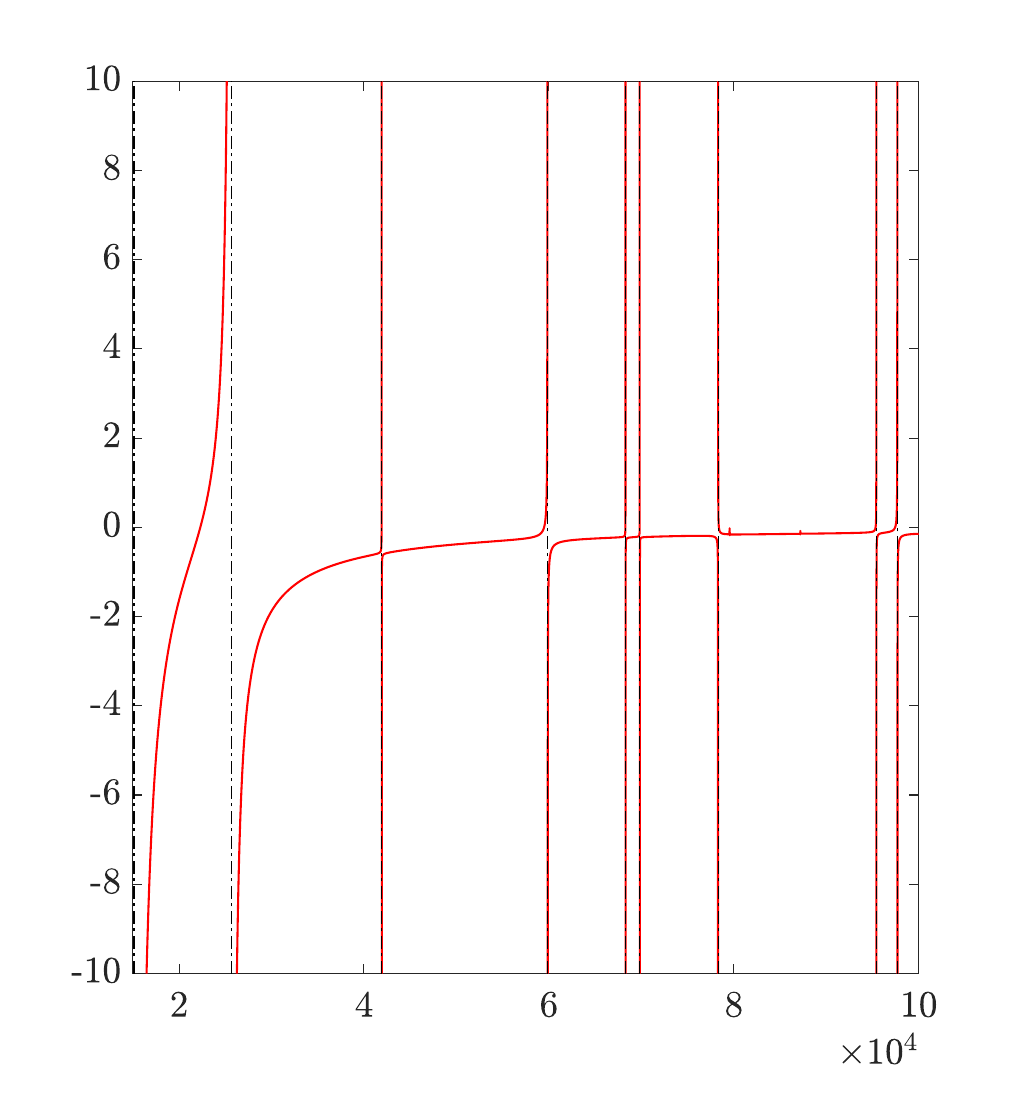}
		\put(-107,190) {Soft}
		\put(-90,0) {${\omega}$}
		\put(-193, 80) {\rotatebox{90}{$\mu_2 u(r^*,z^*)$}}
		\put(-195,190) {{\bf (b)}}
		
		\vspace{2mm}
		
		\includegraphics[width=0.48\textwidth]{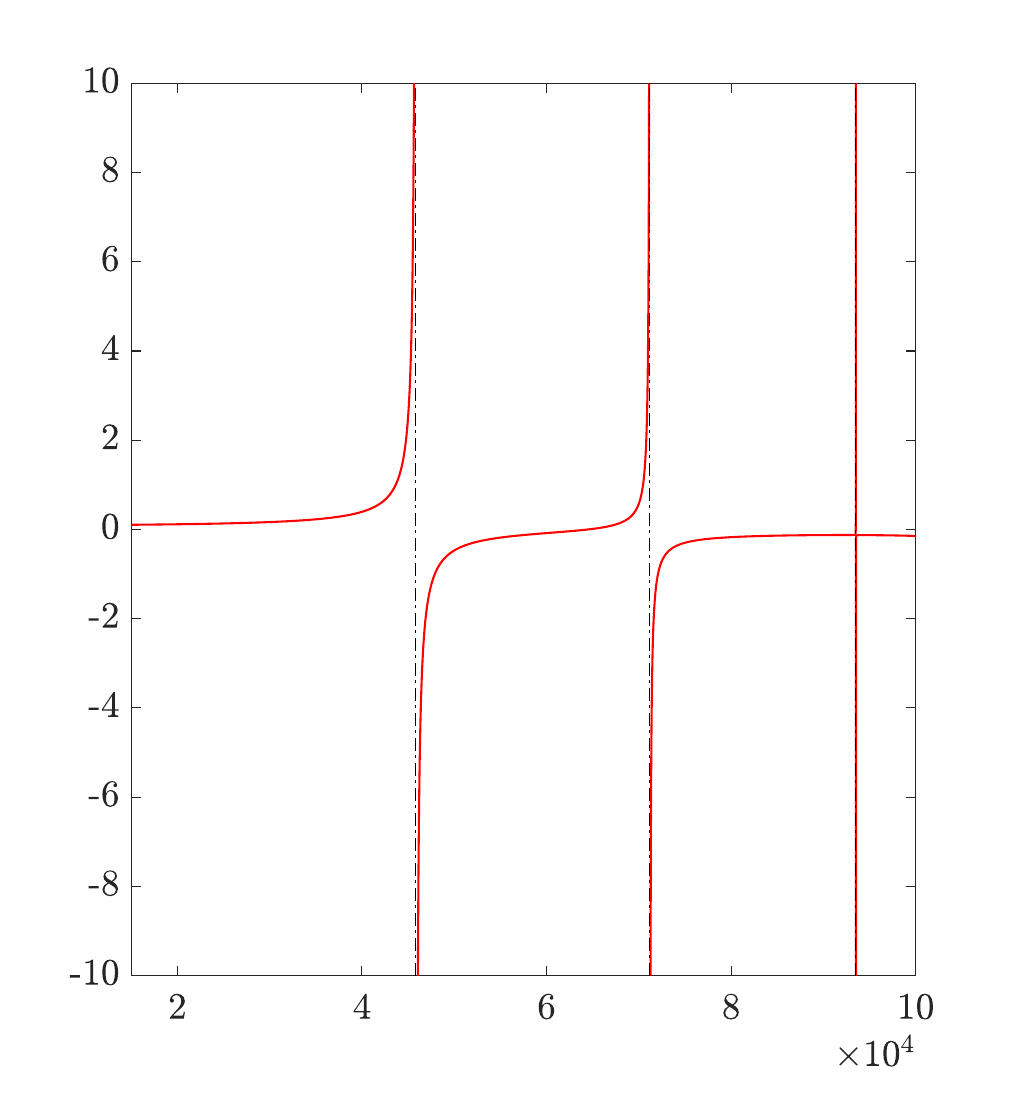}
		\put(-107,190) {Rigid}
		\put(-90,0) {${\omega}$}
		\put(-193, 80) {\rotatebox{90}{$\mu_2 u(r^*,z^*)$}}
		\put(-195,190) {{\bf (c)}}
		\hspace{4mm}
		\begin{minipage}{0.48\textwidth}
			\vspace{-7.5cm} \hspace{1.4cm}
			\begin{tabular}{| c | c | c |}
				\hline
				Ideal & Soft & Rigid \\
				\hline \hline
				4.908 & 1.512 & 4.586 \\
				\hline
				5.240 & 2.568 & 7.120 \\
				\hline
				5.988 & 4.190 & 9.351 \\
				\hline
				7.044 & 5.988 & \\
				\hline
				8.258 & 6.828 & \\
				\hline
				9.537 & 6.984 & \\
				\hline
				& 7.826 & \\
				\hline
				& 9.533 & \\
				\hline
				& 9.740 & \\
				\hline \hline
			\end{tabular}
			\put(-133,-20) {{\small \rotatebox{90}{$ \omega^* \times 10^4$ [Hz]}}}
		\end{minipage}
		\put(-195,190) {{\bf (d)}}
		\caption{{The resonance frequencies computed from the scaled displacement $\mu_2 u$ at the centre $r^*=(a+R)/2, z^*=H/2$ of a hollow cuprum-aluminium cylinder  for the various coupling conditions: (a) ideal, (b) soft, (c) rigid. The red lines provide the computed scaled displacement against the vibration rate ${\omega}$. The location of phase changes, denoted by black lines, indicates the resonance frequencies $\omega^*$. The resonance frequencies $\omega^*$ for each coupling condition are listed in Table (d) to four significant figures. }}
		\label{Fig:Resonance}
	\end{figure}
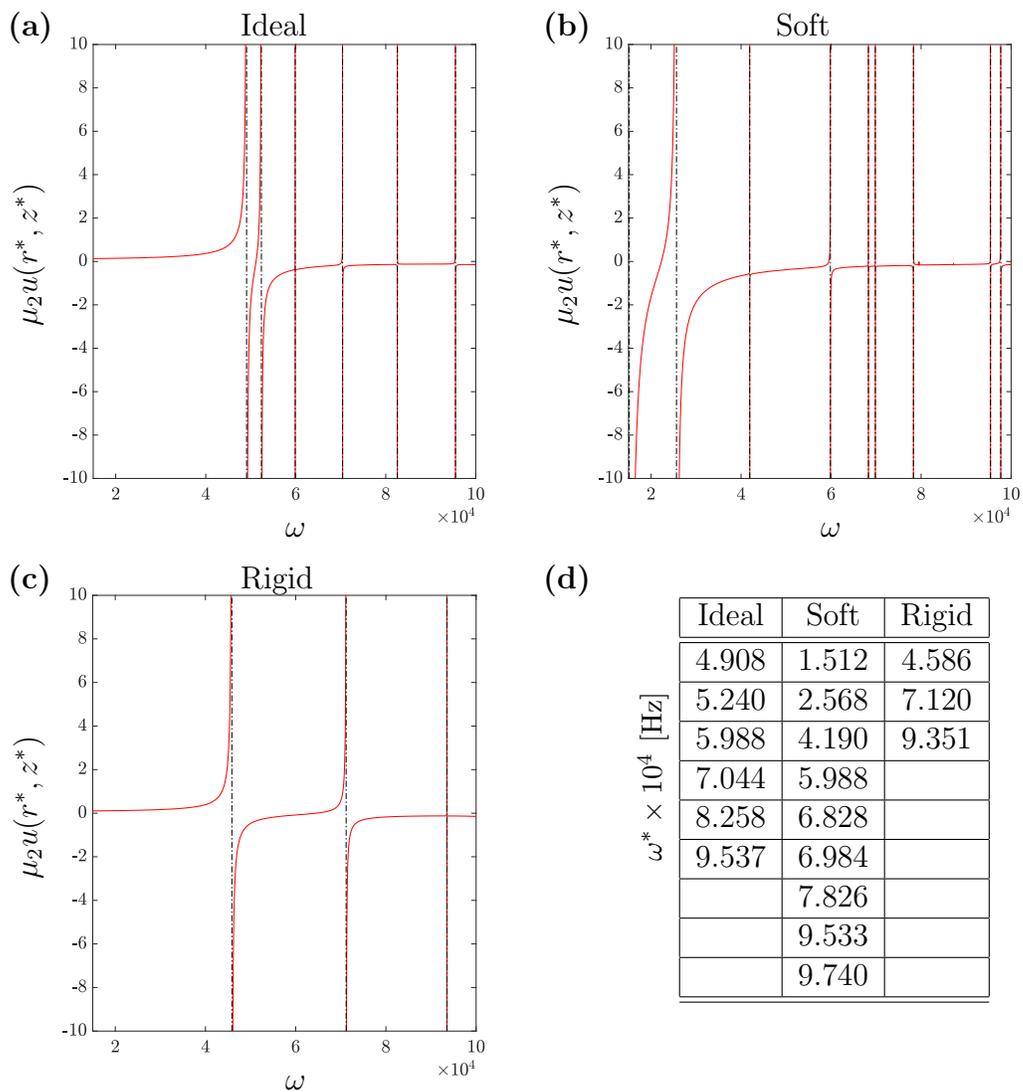
	
	{First, we should note the comparable work in the literature. The initial models for the resonance frequency of finite-length rotating cylinders we provided by \cite{PADOVAN1973469,ZOHAR1973269}. These led to a wider investigation of rotating and vibrating cylinders, with a summary of approaches for different  vibrational modes, alongside comparison with experiments, provided by \cite{WANG1996955}. There has been a corresponding interest in the resonance frequencies for vibrating cylindrical shells, see for example \cite{HOSSEINIHASHEMI20139} and references therein. Finally, an approach for describing hollow cylinders under a wide variety of loading modes was given by \cite{Ebenezer2015}. These modes however assumed that the upper and lower surfaces were subject to identical loading, and the case of torsion was not considered. The presented work therefore provides the first description of a vibrating hollow cylinder under torsion loading, with differing loading on the upper and lower surfaces, and the first for a coupled cylinder. }
	
	{We again consider and cuprum-aluminium cylinder, whose geometry and material parameters are as outlined in Table.~\ref{Table:Material}. The location of resonance frequencies are obtained utilizing the approach outlined in \cite{Grinchenko1981}. {Namely, the value of the stress and displacement at the cylinder centre $r^*=(a+R)/2, z^*=H/2$ is computed for $5000$ uniformly distributed values of the frequency over the range $1.5\times 10^4 \leq \omega \leq 10^5$ [Hz], keeping all other parameters fixed. An increase in the dynamic stress state with $\omega$ indicates that one is approaching a resonance frequency, while the transition through a resonance frequency is identified by a sharp phase shift in the displacement and stress response curves (see Fig.~\ref{Fig:Resonance}a-c). This provides the initial intervals containing the resonance frequencies. An iterative procedure, refining the mesh within each interval and repeating the process, can then by used to determine the corresponding resonance frequencies. The key advantages of this approach are the relative ease and speed at which resonance frequencies can be located,  the high level of accuracy that can be achieved, and that it only relies on having the existing solution data.}
		
		{An obvious disadvantage of this algorithm is that as the resonance frequency is approached the accuracy of the boundary conditions deteriorates. This limits the division process and the degree of accuracy in determining the resonance frequencies. To account for this, the criterion for the error of determined eigenvalues is taken as the accuracy of the boundary conditions' executions. In this section it was taken as $10^{-5}$. This allowed the values of the resonance frequencies to be obtained to four significant figures, which are provided in Fig.~\ref{Fig:Resonance}d.}
		
		
		{It should be noted that this approach is not unique, and other approaches for investigating wave-phase behaviour near resonance frequencies can also be applied. For example, \cite{Postnova2008} utilized similar techniques to those presented here to investigate edge and trapped mode resonances in elastic bodies. Meanwhile, in \cite{Krack2016} a numerical method using phase lag to track resonant curves was applied. This allowed for tracing the locus of a specific peak in the frequency response when varying key system parameters, and was applied to the periodic, steady-state vibrations of harmonically forced nonlinear mechanical systems. Further, \cite{Deng2020} demonstrated that such results have application to classical dynamic elastic problems. Their work investigated nanoparticles via the framework of linear elasticity, demonstrating dependence of the perturbed wave-phase behaviour on the closeness to elastic resonance frequencies. }

	{Let us now return to the results presented in Fig.~\ref{Fig:Resonance}, providing the resonance frequencies for the differing coupling conditions in the centre of a cuprum-aluminium cylinder.}  {It is clear from these that the form of the coupling condition has a significant impact on the number of, and value of, the resonance frequencies of the cylinder. For the ideal coupling condition there exist $6$ resonance frequencies in the consider range of $\omega$, while $9$ exist for the soft case and only $3$ for the rigid interface. Interestingly, there are no shared eigenfrequencies between the cases. Further investigation by the authors show that this is the case throughout the cylinder, rather than being localised to the interface.}
	
	

	\section{Using an interfacial layer to approximate damage within a cylinder}\label{Sect:DamageApprox}
	
	The presented formulation and solutions for the displacement allows for modelling a coupled cylinder with various interfacial coupling conditions. However, it may also be possible to use this to approximate damage within a {single material cylinder at height $z=h$ }, such as the presence of a hidden ring crack. This can be achieved by considering a weak interface at $z=h$, which mimics the reduced bonding between the parts of the cylinder above and below the damaged region. Such an approximation can be created provided some link between the effect of the damage and the shear modulus of the interfacial layer can be obtained.

	{To demonstrate that this is possible, we perform a comparison with  \cite{ZHURAVLOVA2024103976} which considered a ring crack within a single-material hollow cylinder. Let us} consider a small cylinder whose upper and lower sections are made of steel, with some soft interface between them having height $h_0 = 10^{-4}$ [m] (mimicking the ring crack in the stated paper). The cylinder has inner radius $a = 5.5\times 10^{-3}$ [m], outer radius $R = 7.5\times 10^{-3}$ [m], and height $H=5\times 10^{-3}$ [m]. Note that the comparison between the model presented here and the work of \cite{ZHURAVLOVA2024103976} is not perfect, as that paper assumes loading applied to the inner walls of the cylinder $r=a$ while here we assume loading on the upper face of the cylinder $z=H$, however this will not significantly impact the result.
	
	We consider the displacement $u(r,z)$ at the top of the cylinder $z=H$, as that where measurements are most easily taken in practical applications. Let us investigate the ratio of the displacements between the cylinder with a soft interface and the ideal case 
	$
	{u_{\text{soft}}(r,H)} / {u_{\text{ideal}}(r,H)} .
	$
	Note that in practical application this would correspond to the observed displacement divided by that predicted for {undamaged cylinder}. This ratio for various relative interface positions $\delta = h/H$ and shear moduli ratio $\mu_1 / \mu_0$ are provided in Fig.~\ref{Fig_WeakApprox}. It can be seen that the interface position and shear moduli ratio both play a key role in determining the magnitude of the resulting displacement. However, the distribution of that displacement over $r$ is only small compared to the magnitude of the displacement.
	
	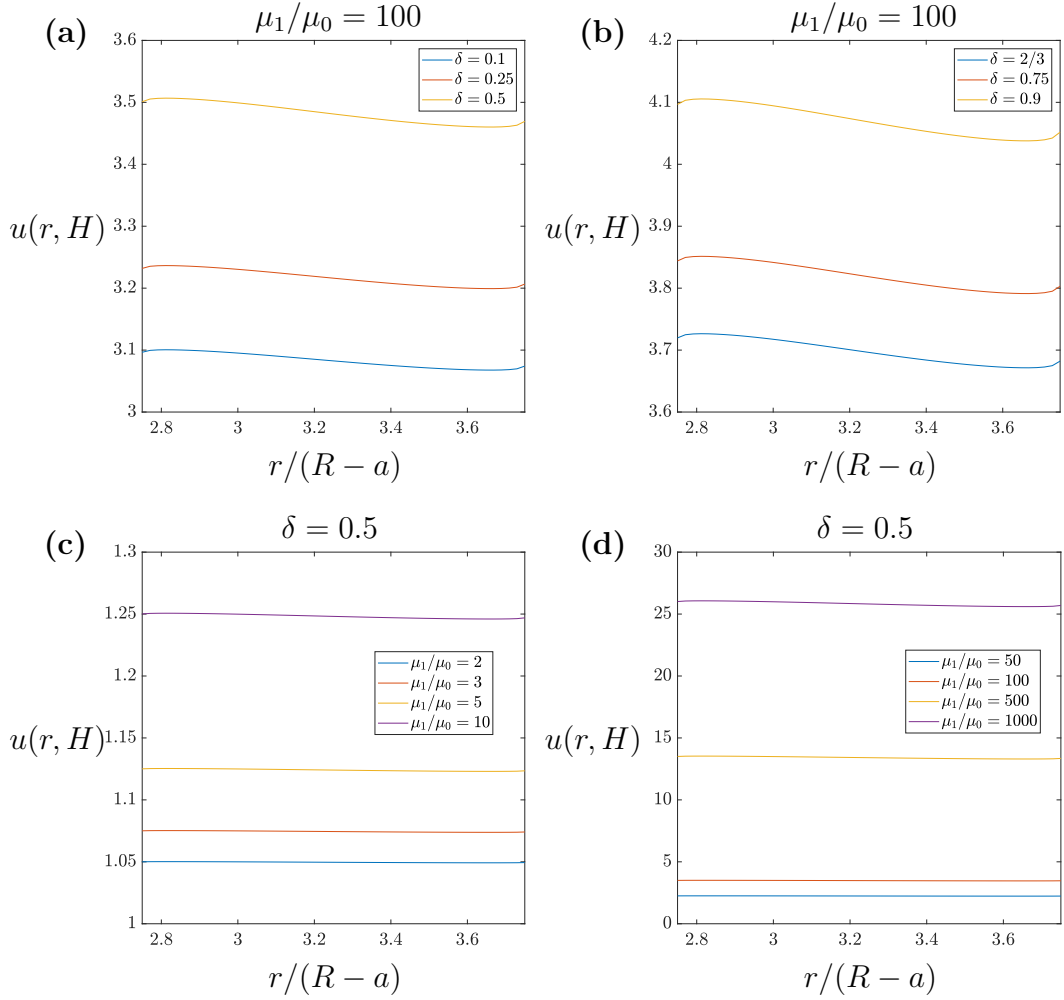
\begin{figure}[t!]
		\centering
		\includegraphics[width=0.48\textwidth]{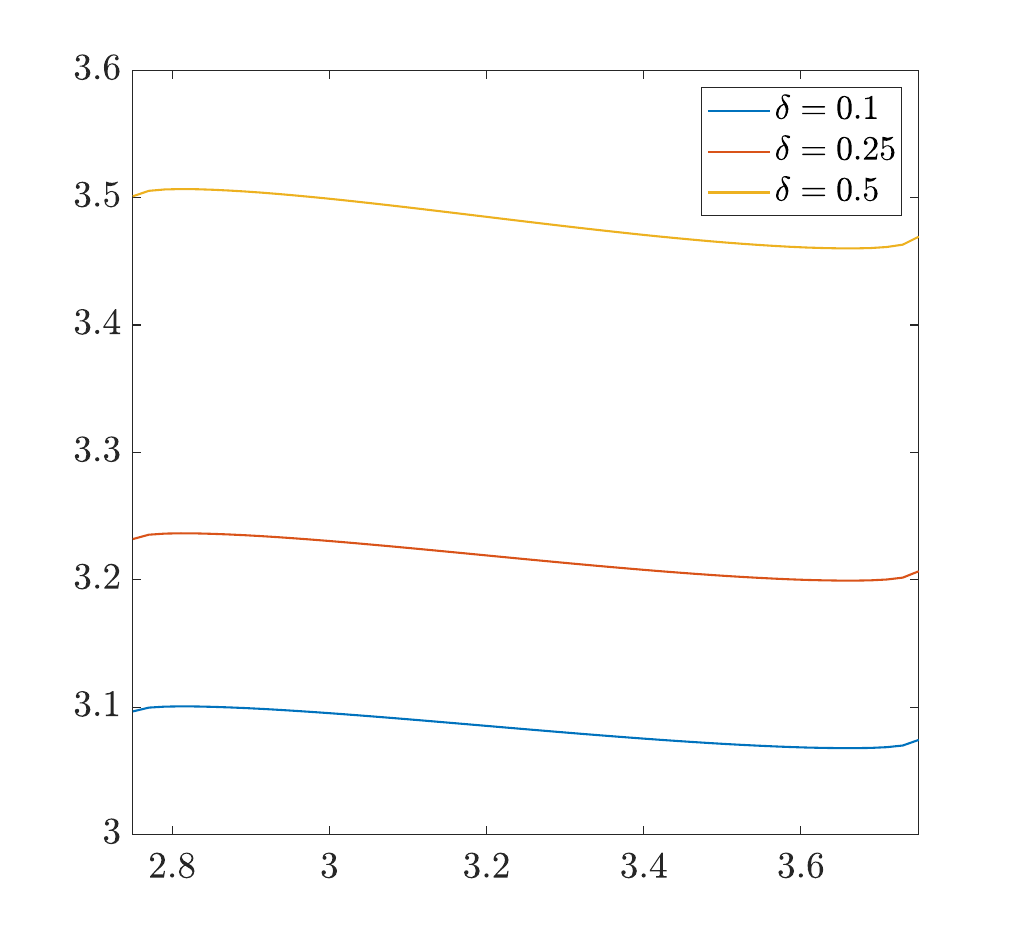}
		\put(-120,165) {$\mu_1 / \mu_0 = 100$}
		\put(-115,-5) {$r/(R-a)$}
		\put(-213, 85) {$u(r,H)$}
		\put(-200,159) {{\bf (a)}}
		\hspace{4mm}
		\includegraphics[width=0.48\textwidth]{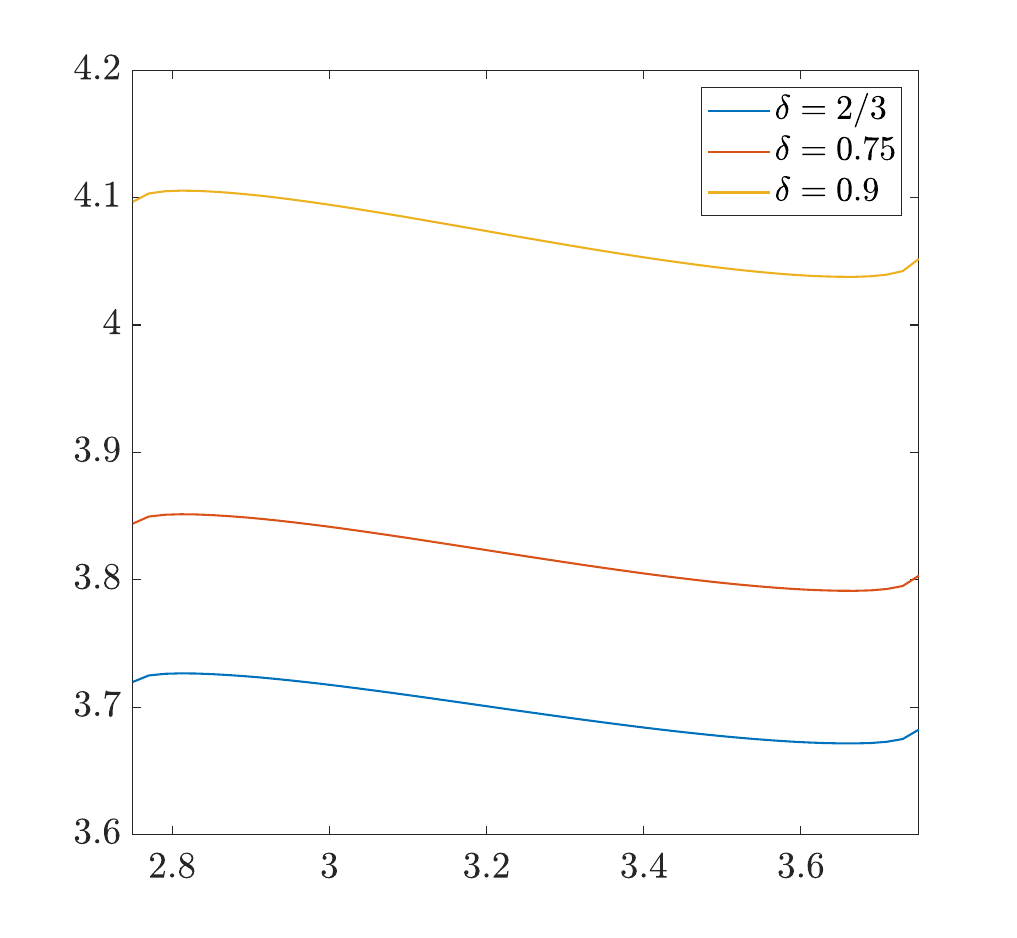}
		\put(-120,165) {$\mu_1 / \mu_0 = 100$}
		\put(-115,-5) {$r/(R-a)$}
		\put(-213, 85) {$u(r,H)$}
		\put(-200,159) {{\bf (b)}}
		
		\vspace{4mm}
		
		\includegraphics[width=0.48\textwidth]{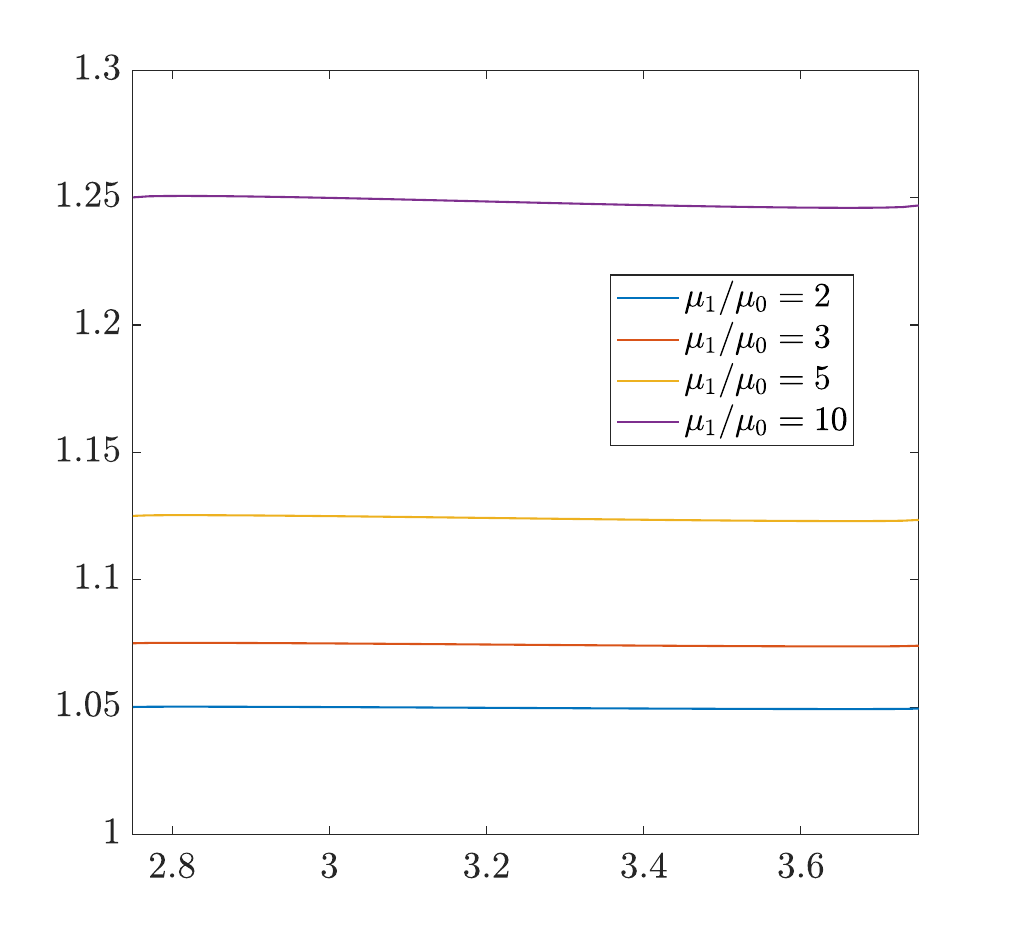}
		\put(-110,165) {$\delta = 0.5$}
		\put(-115,-5) {$r/(R-a)$}
		\put(-213, 85) {$u(r,H)$}
		\put(-200,159) {{\bf (c)}}
		\hspace{4mm}
		\includegraphics[width=0.48\textwidth]{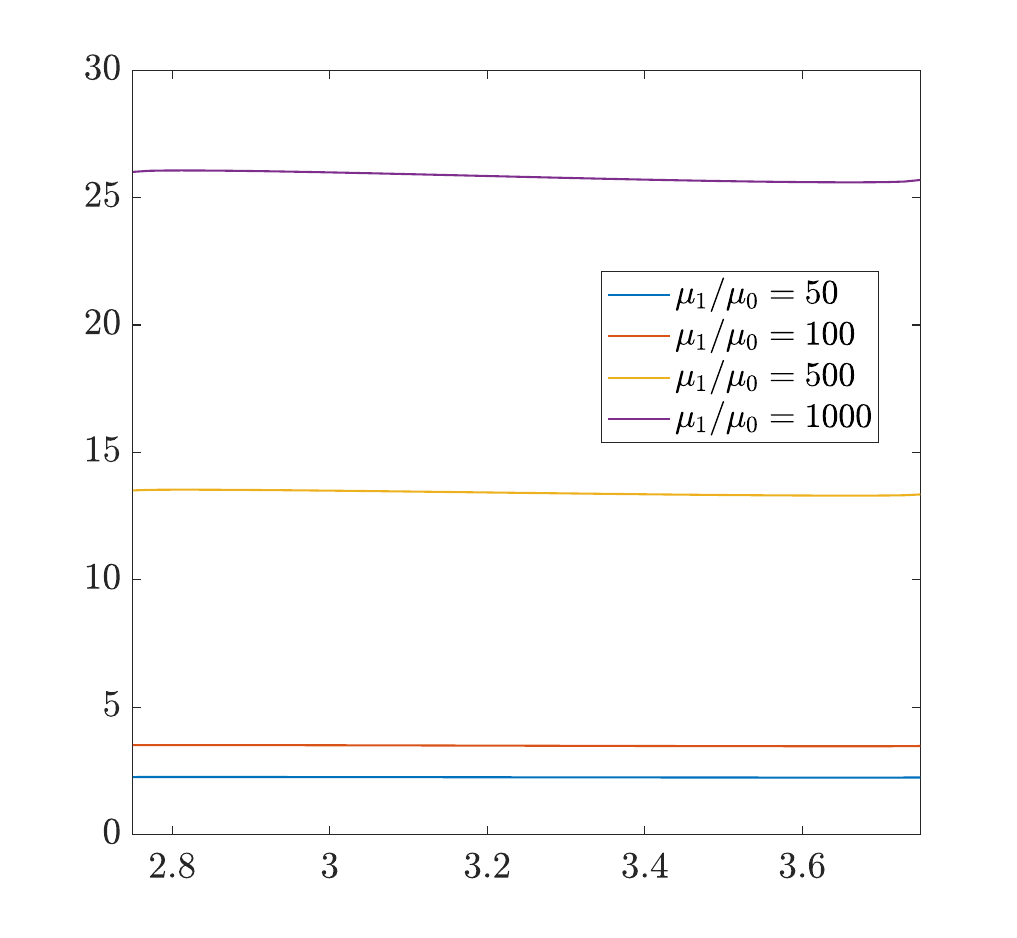}
		\put(-110,165) {$\delta = 0.5$}
		\put(-115,-5) {$r/(R-a)$}
		\put(-213, 85) {$u(r,H)$}
		\put(-200,159) {{\bf (d)}}
		\caption{The displacement at the top of the cylinder $u(r,H)$ for: (a), (b) fixed shear moduli ratio $\mu_1 / \mu_0 = 100$ and varying relative interface height $\delta = h/H$; (c), (d) fixed relative interface height $\delta = 0.5$ and varying shear moduli ratio $\mu_1 / \mu_0$.}
		\label{Fig_WeakApprox}
	\end{figure}
	
	Consequently, it makes sense to consider the arithmetic mean of the displacement ratio over radial points $r=[r_1,\hdots ,r_N]$ which, noting \eqref{ideal_displacement}, \eqref{soft_displacement}, the formulae for the Green's function \eqref{Greens_Func}, \eqref{Greens_Func0}, the definition of $K_1$, and that $\mu_1 \equiv \mu_2$, $c_1 \equiv c_2$ (so $\beta_1 \equiv \beta_2$, $\gamma_{1n}\equiv \gamma_{2n}$), is given by
	\begin{equation}\label{arithmetic_ratio}
		\begin{aligned}
			\bar{u} =& \frac{1}{N}\sum_{i=1}^N \frac{u_{\text{soft}}(r_i, H)}{u_{\text{ideal}}(r_i, H)} \\
			=& 1 + \frac{\mu_1 h_0}{\mu_0 N}\sum_{i=1}^N \frac{\frac{4r_i p_0}{R^4-a^4} \frac{\cos(\beta_2 h)}{\cos(\beta_2 H)} \frac{\cos(\beta_2 h)}{\Delta_{20}^S}-\sum_{n=1}^\infty p_n  \frac{\cosh(\gamma_{2n}h)}{\cosh(\gamma_{2n}H)} \frac{\cosh(\gamma_{2n} h)}{\Delta_{2n}^S} \frac{K(\lambda_n, r_i)}{\| K(\lambda_n , r_i)\|^2} }{\frac{4r_i p_0}{\beta_2 (R^4 - a^2)} \tan(\beta_2 H) + \sum_{n=1}^\infty \frac{p_n}{\gamma_{2n}} \tanh (\gamma_{2n} H)\frac{K(\lambda_n , r_i)}{\| K(\lambda_n , r_i)\|^2} },
		\end{aligned}
	\end{equation} 
	where in this special case $\mu_1\equiv\mu_2$, $c_1 \equiv c_2$
	$$
	\Delta^S_{20} = \cos(\beta_2 H) + \frac{\mu_1 h_0}{\mu_0} \beta_2 \sin\left[\beta_2 (H-h)\right] \cos(\beta_2 h), $$ $$ \Delta^S_{2n} = \cosh(\gamma_{2n} H) + \frac{\mu_1 h_0}{\mu_0} \gamma_{2n} \sinh\left[\gamma_{2n} (H-h)\right] \cosh(\gamma_{2n} h).
	$$
	This expression indicates an almost linear relationship between the shear modulus ratio $\mu_1/\mu_0$ and the arithmetic mean displacement ratio $\bar{u}$. This is confirmed in Fig.~\ref{Fig_Ubar}, with least-squares analysis revealing that for the given material parameters and cylinder geometry $\bar{u}$ can be accurately described by (to $4$ significant figures)
	\[
	\bar{u} = 1.000 + 0.02482 \frac{\mu_1}{\mu_0}.
	\]
	In fact, this linear approximation yields a relative error below $0.01\%$ for all of the points computed, which is of comparable order to the accuracy of computations. It should be noted however that the weak interface assumptions mean that the error will be slightly higher in the limit $\mu_1 / \mu_0 \to 1$ as the right-hand side of \eqref{arithmetic_ratio} doesn't go to zero (we have $\bar{u}\to 1.02$ rather than $\bar{u} \to 1.00$, to three significant figures). However, given that we would assume $\mu_0 < \mu_1$ in the case of a damaged cylinder, this means that the linear approximation can be used in all relevant cases.
	
	\begin{figure}[tp!]
		\centering
		\includegraphics[width=0.48\textwidth]{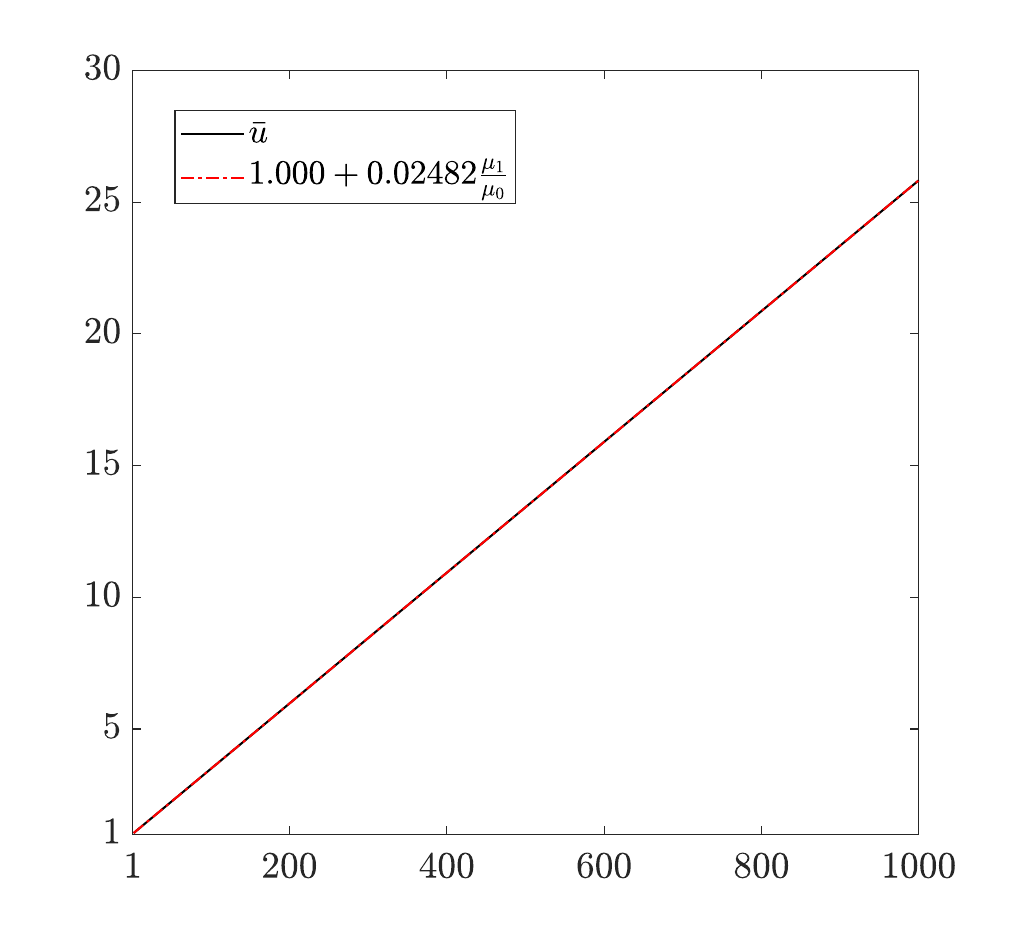}
		\put(-103,-5) {$\mu_1 / \mu_0$}
		\put(-190, 85) {$\bar{u}$}
		\caption{The arithmetic mean over $r$ of the displacement ratio $u_{\text{soft}}/u_{\text{ideal}}$ \eqref{arithmetic_ratio} at the interface $z=h$, alongside a linear approximation produced using the least-square method. Results are for a steel cylinder with fixed interface relative height $\delta = h/H = 0.5$.}
		\label{Fig_Ubar}
	\end{figure}
	
	{This provides a practical approach for engineers to estimate internal damage by comparing observed displacements with those predicted by the ideal model. Namely, the interface shear modulus $\mu_0$ for approximate the damaged cylinder is chosen based on the known magnitude} of the arithmetic mean of the displacement ratio $\bar{u}$, the location at which the damage occurs $h$, and the height of the damaged region $h_0$, based on a linear approximation of \eqref{arithmetic_ratio}. {The weak interface model, taking these values of $\mu_0$ and $h, h_0$, can then be used to estimate the stress and displacement within the cylinder.}

	One additional benefit of the weak interface approximation comes from a crucial difference between the displacement observed for the weak interface and that when a ring crack is present. Namely, the profile over $r$ of the displacement ratio compared to the ideal case $u_{\text{soft}}(r,h)/u_{\text{ideal}}(r,h)$. As seen in Fig.~\ref{Fig_WeakApprox}, the weak interface provides an almost constant displacement ratio over $r$, due to the impact on the bonding between the upper and lower cylinders being equally spread across the entire interface. {However, in \cite{ZHURAVLOVA2024103976} it} was demonstrated that a ring crack with produce significant variation in the displacement over $r$ - most notably the clear presence of a maximum (see Fig.~4 in that paper). Therefore, the weak interface approximation allows for a simple method of approximating the relative magnitude of the displacement, while the distribution of the displacement across the cylinder provides a non-destructive test to infer the type of damage present (e.g. whether or not a crack is forming) and the proportion of the `interface' that it covers.
	
	\section{Summary and concluding remarks}\label{Sect:Conclusion}
	
	The torsion problem of a coupled hollow cylinder subject to various forms of coupling between the upper and lower cylinders' (ideal, soft, rigid) was considered. The cylinders were of arbitrary geometry and torsion loading on the upper face, while the bottom was fixed in place and the cylinder' sides assumed free from loading. The solution for the displacement and tangential stress in each case were obtained in terms of the Green's function, utilizing the finite Hankel transform. {Many aspects of the model are unique, being the first to consider vibration of a (vertically) coupled cylinder, to account for a multitude of coupling conditions between upper and lower sections, and to consider torsional loading in addition to the vibration. Model results also allow the impact of coupling condition type on the resonance frequencies of the cylinder to be investigated, and provide a novel approach to approximating damage within a cylindrical structure, amongst other new possibilities.}
	
	The results of simulations demonstrated a number of key trends:
	\begin{itemize}
		\item There is a close correspondence between the results for the ideal and rigid interface, although there is some difference in behaviour near the bottom of the cylinder (see Fig.~\ref{Fig_Layers_Mid}a,b). Between these two cases, the relative difference in the displacement at the interface remained below $0.6$\% for the long thin cylinder ($R/H=0.5$) for all shear moduli ratios $\mu_0 / \mu_1$, and below $7$\% for the modest shear ratio $\mu_0/\mu_1 = 10$, but could rise as high as $24$\% for short fat cylinders ($R/H\geq 2.5$) with extreme shear moduli ratios (see Table.~\ref{RigidDiff}). Meanwhile, the relative difference in the tangential stresses was $<5$\%, for all simulated shear moduli and cylinder geometries. 
		\item There is a clear difference in behaviour between the soft interface and that for ideal or rigid contact, with a relative difference of the displacement at the interface between $65$\% - $4667$\%, depending on the cylinder geometry and interface material, although the relative difference in the tangential stress is far smaller (see Table.~\ref{SoftDiff}). There is however a close correspondence in the displacement and stress between the soft and rigid interfaces at the bottom of the cylinder (see Fig.~\ref{Fig_Layers_Mid}a,b).
		\item {The resonance frequencies for the cylinder can be obtained using the presented model. The presented method is highly robust allowing one to obtain the resonance frequencies to arbitrarily high levels of accuracy, making it a useful tool for bifurcation analysis for both coupled and homogeneous cylinders (via assumption of an ideal interface). It was shown that the resonance frequency varied considerable depending on the form of the coupling condition (see Fig.~\ref{Fig:Resonance}).}
		\item The impact of {vibration} frequency $\omega$ on the cylinder displacement and tangential stress {is localised to the region about the eigenfrequencies, although the size of this region varies significantly between eigenfrequencies (see Fig.~\ref{Fig:Resonance}). Away from these regions the vibration frequency only has a small effect compared to that of changing the shear modulus and cylinder geometry, and has a negligible effect on the distribution over $r$}. The {impact} is however far more significant for the case of a soft contact condition than for the ideal or rigid case (see Fig.~\ref{Fig_vibration}).
		\item The impact of the particular torsion loading on the displacement  depends primarily upon the magnitude of the torsion loading, with some weighting towards loads applied towards the outer cylinder edge $r=R$ (see the form and impact of $p_n$ \eqref{transformed_problem} on the displacement \eqref{ideal_displacement}, \eqref{soft_displacement},  \eqref{rigid_displacement}).
	\end{itemize} 
	
	The results for the stress and displacement obtained in this work can be directly applied when considering the construction of coupled cylinders in {application, such as in turbines, the automotive industry, or design of biomedical components. For example, the ability to consider differing coupling conditions can be used to determine optimal adhesive properties between cylinder segments.} 
	
	The presented model may also be used to approximate damage within a single {cylinder, providing updated displacement predictions that can inform risk assessments and performance estimates}. This is achieved by approximating the damaged region using a weak interface, with the method for doing so being outlined in Sect.~\ref{Sect:DamageApprox}. If some mean difference in the displacement ratio $\bar{u}$ (between the ideal/ `expected' displacement and that measured) at the top of the cylinder is detected, and the damage is believed to start at $z=h$ and have height $h_0$, then a linear approximation of \eqref{arithmetic_ratio} can be used to determine the appropriate interface shear modulus $\mu_0$ for the approximation. The displacement then follows easily from \eqref{soft_displacement}. Further, the variation of the displacement over $r$ measured at the top of the cylinder can be used to infer the type of damage present, as it is directly linked to the extent of bonding between the regions above and below the damaged area{, providing a potential avenue for use in non-destructive testing}. If there is little variation over $r$ then there is likely almost uniform damage over the damaged area, while the presence of notable maxima may indicate a ring crack (see \cite{ZHURAVLOVA2024103976}, Fig.~4).
	
	It should be noted that throughout this work we have assumed the {vibrational} frequency of the cylinder $\omega\neq 0$. When considering the case where the frequency $\omega=0$, the systems governing equation reduces to the Laplacian $\Delta u_k = 0$ within each cylinder $k=1,2$. The model presented in this work can therefore be used to describe a range of other problems, such as the heat flux across a pair of joined cylinders, with applications to geothermal wells, amongst others. This could be further improved through the inclusion of thermoelastic effects within the model, in particular with regards to the interface. This will be the subject of future work by the authors. 
	
	\section*{CRediT authorship contribution statement}
	
	{\bf Igor Istenes} initiated the research, contributed ideas, analysis of results. {\bf Daniel Peck} analysis of results, visualisation, produced the manuscript. {\bf Yuriy Protserov} contributed ideas, analysis of results. {\bf Natalya Vaysfeld} contributed ideas, analysis of results, reviewed/edited the manuscript. {\bf Zinaida Zhuravlova} performed the derivations, initial visualisation, analysis of results, produced the code, produced the manuscript, oversaw all research.
	
	\section*{Funding}
	All authors were supported under European project funded by Horizon 2020 Framework Programme for Research and Innovation
	(2014–2020) (H2020-MSCA-RISE-2020) Grant Agreement number 101008140 EffectFact ``Effective Factorisation techniques for
	matrix-functions: Developing theory, numerical methods and impactful applications’’. D. Peck acknowledge supports from a project within the Innovate Ukraine competition, funded by the UK International Development and hosted by the British Embassy Kyiv. N. Vaysfeld acknowledges support from the Royal Society Wolfson Visiting Fellowship R3/233003.
	
	\section*{Acknowledgements}
	The authors are exceptionally grateful to the reviewers {and IJES editor for their helpful comments, alongside Prof. G. Mishuris for his critical insights when formulating and realizing this work.}

	\bibliography{Penny_Bib}
	\bibliographystyle{apalike}

\end{document}